\documentclass[12pt,preprint,amssymb]{aastex}

\usepackage{graphicx}
\usepackage[figuresright]{rotating}

\def\asca{{\em ASCA}\/}

\def\HI{H~{\sc i}}
\def\Ha{H$\alpha$}

\shorttitle{Probe Galaxy-ICM Interaction}
\shortauthors{Gu et al.}

\begin{document}
\title{Galaxy Infall by Interacting with its Environment: \\ a Comprehensive Study of 340 Galaxy Clusters}

\author {Liyi Gu\altaffilmark{1,2,3}, Zhonglue Wen\altaffilmark{4}, Poshak Gandhi\altaffilmark{5,6}, Naohisa Inada\altaffilmark{7}, Madoka Kawaharada\altaffilmark{8}, Tadayuki Kodama\altaffilmark{9}, Saori Konami\altaffilmark{10}, Kazuhiro Nakazawa\altaffilmark{2}, Haiguang Xu\altaffilmark{11,12}, and Kazuo Makishima\altaffilmark{2,3,13}}

\altaffiltext{1}{SRON Netherlands Institute for Space Research, Sorbonnelaan 2, 3584 CA Utrecht, the Netherlands}
\altaffiltext{2}{Department of Physics, University of Tokyo, 7-3-1
  Hongo, Bunkyo-ku, Tokyo 113-0033, Japan}
\altaffiltext{3}{Research Center for the Early Universe, School of Science, University of Tokyo, 7-3-1 Hongo, Bunkyo-ku, Tokyo 113-0033, Japan}
\altaffiltext{4}{National Astronomical Observatories, Chinese Academy of Sciences, 20A Datun Road, Chaoyang District, Beijing 100012, China}
\altaffiltext{5}{School of Physics \& Astronomy, University of Southampton, Highfield, Southampton SO17 1BJ, UK}
\altaffiltext{6}{Department of Physics, Durham University, South Road, Durham DH1 3LE, UK}
\altaffiltext{7}{Department of Physics, Nara National College of Technology, Yamatokohriyama, Nara 639-1080, Japan}
\altaffiltext{8}{Institute of Space and Astronautical Science, Japan Aerospace Exploration Agency, 3-1-1 Yoshinodai, Chuo-ku, Sagamihara, Kanagawa 229-8510, Japan}
\altaffiltext{9}{National Astronomical Observatory of Japan, Mitaka, Tokyo 181-8588, Japan}
\altaffiltext{10}{Department of Physics, Tokyo Metropolitan University, 1-1 Minami-Osawa, Hachioji, Tokyo 192-0397}
\altaffiltext{11}{Department of Physics and Astronomy, Shanghai Jiao Tong University, 800 Dongchuan Road, Minhang, Shanghai 200240, China}
\altaffiltext{12}{IFSA Collaborative Innovation Center, Shanghai Jiao Tong University, 800 Dongchuan Road, Minhang, Shanghai 200240, China}
\altaffiltext{13}{MAXI Team, Institute of Physical and Chemical Research, 2-1 Hirosawa, Wako, Saitama 351-0198, Japan}

\begin{abstract}

To study systematically the evolution on the angular extents of the galaxy, ICM, and dark matter 
components in galaxy clusters, we
compiled the optical and X-ray properties of a sample of 340 clusters with redshifts $<0.5$, based on all the available data with
the Sloan Digital Sky Survey (SDSS) and {\it Chandra}/{\it XMM-Newton}. For each cluster, the member galaxies
were determined primarily with photometric redshift measurements. The radial ICM mass distribution, as well as the
total gravitational mass distribution, were derived from a spatially-resolved spectral analysis of the X-ray data.
When normalizing the radial profile of galaxy number to that of the ICM mass, the relative curve was found to depend significantly on the 
cluster redshift; it drops more steeply towards outside in lower redshift subsamples. The same evolution is
found in the galaxy-to-total mass profile, while the ICM-to-total mass profile varies in an opposite way. 
The behavior of the galaxy-to-ICM distribution does not depend on the cluster mass, suggesting that the detected redshift-dependence
is not due to mass-related effects, such as sample selection bias. Also, it cannot be ascribed to various redshift-dependent 
systematic errors. We interpret that
the galaxies, the ICM, and the dark matter components had similar angular distributions when a cluster was formed, while
the galaxies travelling interior of the cluster have continuously fallen towards the center relative to the other
components, and the ICM has slightly expanded
relative to the dark matter although it suffers strong radiative loss. This cosmological galaxy infall, accompanied by an ICM 
expansion, can be explained by considering that the galaxies interact strongly with the ICM while they are moving 
through it. The interaction is considered to create a large 
energy flow of $10^{44-45}$ erg $\rm s^{-1}$ per cluster from the member galaxies to their environment, which is expected to 
continue over cosmological time scales.

\end{abstract}

\section{INTRODUCTION}

In the standard cosmological model, the formation of large-scale structure is dominated by gravitational 
dynamics, while the gas physics plays a minor role. The gravitational collapse of cosmic matter over 
several megaparsecs creates galaxy clusters, the largest virialized system in the Universe. On small-scale
domains (e.g, galaxies), instead, non-gravitational processes, related to the dynamics and evolution of baryons,
become more and more important. The physical states of galaxies and intracluster medium (ICM) are 
shaped by complex processes such as radiative cooling, feedback from supernovae and active galactic nuclei, star formation, and
interactions between galaxies and ICM. Because galaxy clusters stand at the transition between the two scale domains, 
they are often studied for both cosmological and astrophysical aims. It is hence crucial to quantify how
the astrophysical processes have affected the cosmological properties of the cluster baryons.

So far, the cluster member galaxies and the ICM have been assumed in many cases to be subject to different sets
of astrophysical processes, and evolve separately over
cosmological time scales. However, based on X-ray observations of the ICM with {\it ASCA}, Makishima et al. (2001; M01 hereafter)
proposed a novel picture: physical interactions occur universally between member galaxies and the ICM, which transfer free 
energy from galaxies to the ICM, and drag the galaxies to ultimately fall to the cluster center over the Hubble time.  
Indeed, possible remnants of the interaction have been observed around many galaxies in \HI\ (e.g., Oosterloo \& van Gorkom 2005),
\Ha\ (e.g., Yoshida et al. 2008), and X-ray bands (e.g., Sun et al. 2006; Gu et al. 2013b).

This simple idea can potentially shed light on several unsolved issues. First, it immediately explains why
the ICM in nearby clusters has a much more extended angular distribution than the member galaxies. The implied member galaxy infall may also
be intimately linked to the formation of central brightest cluster galaxies (or the cD galaxies), as well as that
of diffuse intracluster light. At the same time, it can potentially answer to a long-standing question: 
why galaxies evolve differently in- and out-side clusters? The fraction of blue galaxies, which are considered to be gas-rich 
and star forming objects, increases with redshift
up to $z \sim 0.4$ in the cores of rich clusters (known as Butcher-Oemler effect; Butcher \& Oemler 1984); such a feature is not 
found in low-density environments. This kind of ``environmental effects'', of which the ultimate driving force remained unknown,
may be understood as a consequence of the proposed interactions of the moving galaxies with the ICM. As recently reported in 
Stroe et al. (2015), star formation in member galaxies 
appears to be excited by cluster-scale merging shocks in the ICM. The massive star formation would rapidly consume the molecular
content in the galaxies, and eventually transforms them to the red population. The M01 scenario can also explain, in a natural way,
how the large amount of metals, which must have originally been synthesized in galaxies, are presently distributed to much 
larger radii than the galaxies (e.g., Kawaharada et al. 2009; Matsushita et al. 2013), and the ICM in nearby clusters is metal-enriched 
uniformly up to the periphery (e.g., Werner et al. 2013); the galaxies used to be distributed to larger radii, and enriched that
portion of the ICM while they gradually fall to the center.

Another important consequence expected from the M01 scenario is the energy transfer towards ICM. The high ICM density in cluster 
center would result in a runaway cooling, which leads to the formation of 
enormous cooling flow of gas, and massive star formation in the cD galaxy. However, broad-band X-ray spectroscopy, starting
with {\it ASCA}, found that the effect of cooling is much weaker than previously predicted (e.g., M01; Peterson et al. 2001),
suggesting that some heating mechanisms are in operation. In the galaxy infall scenario, the energy flow from galaxies provides
an important and inherent heating source for the ICM, to be operating essentially in all clusters.

To verify the M01 scenario, the key is to compare the spatial extents of member galaxies and the ICM in clusters at 
different ages. As reported in Gu et al. (2013a; Paper I hereafter), we studied the expected galaxy infall phenomenon using
a statistical sample of 34 massive clusters with redshift range of $0.1-0.9$. We detected, for the first time, a significant
evolution spanning a time interval of $\sim 6$ Gyr in the relative spatial distributions of the cluster galaxies and the ICM;
while the galaxy component was as spatially extended as the ICM at $z > 0.5$, towards the lower redshifts, it has indeed
become more centrally-concentrated relative to the ICM sphere. Since the concentration was found to be rather independent 
of the galaxy mass, it cannot be explained by pure gravitational drag. This result provides an important support
to the galaxy-ICM interaction scenario proposed by M01.

Although the results in paper I are quite firm, the implied view is so novel that further efforts are still needed to make
them more convincing and detailed. To minimize the statistical uncertainty (currently $\sim 20$\%; Fig. 8 of paper I),
it is necessary to increase significantly the sample size. To strengthen the detection of the evolutionary effects, our 
new study should also include
the redshift-dependent angular distributions of the baryon versus dark matter components, which was only briefly studied
in paper I. In addition, we would like to examine in details how the relative distributions of member galaxy, ICM, and 
dark matter components evolve as a function of such parameters as the cluster mass, dynamical state, ICM density, galaxy mass, and galaxy color.  
Such a comprehensive study will enable us to establish conclusively the view of galaxy infall proposed in M01.

We present the new study in the present paper, which has the layout as follows. Section 2 gives a brief
description of the sample selection and data
reduction procedure. The data analysis and results are described in \S 3. We discuss the
physical implication of our results in \S 4, and summarize our work in \S5. Throughout the
paper, we assume a Hubble constant of $H_0=70$$h_{71}$ km s$^{-1}$
Mpc$^{-1}$, a flat universe with the cosmological parameters of
$\Omega_M=0.27$ and $\Omega_\Lambda=0.73$, and quote errors by the 68\%
confidence level unless stated otherwise. The optical magnitudes used
in this paper are all given in the AB system.

\section{OBSERVATION AND DATA PREPARATION}

\subsection{Sample Selection}

To correlate the ICM and galaxy evolution, it is important to construct a large statistical sample of X-ray bright clusters, which 
is based on high quality X-ray and optical data, and has a large coverage in the cluster mass and redshift. Usually the sample size is 
limited by the availability of X-ray data, which is often much less complete than the optical ones. Therefore we started with
selecting all clusters available in the {\it Chandra} and {\it XMM-Newton} archives till 2015, 509 and 442, respectively. The
two archives have an overlap of 262, so that the total number available in X-ray is 689. For the 262 clusters in both X-ray
archives, we selected the data with better signal-to-background ratio (see \S3.5.2 for definition). 
Then, as shown in Table 1, about two-thirds of the X-ray clusters, 468, were found to be covered by the Sloan Digital Sky Survey 
(SDSS) up to the data release 12. The list, consisting of 468 clusters, forms our ``Preliminary Sample''.

We have applied two basic filters to our Preliminary Sample. First, the nearby Virgo, Coma, and Perseus clusters
were discarded for their oversized angular extents. Second, it was further screened to remove those with poor X-ray data quality (i.e., net count $< 5000$).
The remaining number is 381, to be called ``Intermediate Sample''. They were further categorized into three subsamples by their redshift, i.e., 
low-redshift subsample with 138 objects ($z \sim 0.0-0.08$, hereafter
subsample L), intermediate-redshift subsample with 130 objects ($z \sim 0.08-0.22$, hereafter subsample M), and high-redshift subsample consisting of 113 clusters ($z \sim 0.22-0.45$, 
hereafter subsample H). The superposed optical and X-ray images of six example clusters, two for each subsample, are shown in
Figure 1.

\subsection{Completeness Check}

As described above, the sample selection is primarily based on observational limitations, e.g., archival state and X-ray data quality,
rather than on some objective criteria (e.g., flux- or redshift- limited). Therefore, our Intermediate Sample may be 
subject to some selection biases. To examine this issue, we investigate in detail the sample completeness. In essence, minor incompleteness
is acceptable for our purpose, since we already found in Paper I that the major aim of our study, the systematic correlation between
X-ray and optical structures of the clusters, does not depend strongly on the cluster richness.

To test X-ray completeness of the Intermediate Sample, we compared it with existing flux-limited catalogs of the NORAS and REFLEX surveys, which are
known to be fairly complete ($\sim 80$\% and $> 90$\%, respectively; B$\ddot{\rm o}$hringer et al. 2000, 2004). In Figure 2a, the fraction of NORAS/REFLEX
clusters covered by our sample are plotted as a function of the $0.1-2.4$ keV X-ray flux. Our sample recovers $\sim 90$\%/73\% of the sky
density of clusters compared to the NORAS/REFLEX survey at 5$\times 10^{-12}$ ergs s$^{-1}$. Even at 1$\times 10^{-12}$ ergs s$^{-1}$, the relative
completeness reaches 60\%. This means that most of the X-ray detected clusters in the SDSS sky coverage are included in our sample.

To characterize the sample in another way, we calculated the X-ray luminosity functions of sample clusters, normalized to the 
effective survey volume, i.e., a sphere centred on Earth and extending up to the most distant object. Here, the X-ray fluxes were taken from
the NORAS/REFLEX catalogs.
As shown in Figure 2b, the obtained luminosity function is consistent with that of the REFLEX II survey (B$\ddot{\rm o}$hringer et al. 2014) for bright objects, i.e., 
$L_{\rm X} > 10^{44}$ erg/s, while at the fainter end, our Intermediate Sample is less complete, by $\sim 30-40$\%, than the REFLEX II catalog.

A more vital test for this study is to investigate the sample coverage in the cluster mass vs. redshift space. In Figure 2c, we plot 
$M_{500} - z$ relation of our sample clusters, where $M_{500}$ is the total gravitating mass within $R_{500}$, the radius corresponding
to a density contrast of 500 above the cosmological critical value. We measured $M_{500}$ by using the X-ray hydrostatic method described in \S3.3. Most of the 
clusters have $M_{500} = (0.1 - 2) \times 10^{15}$ $M_{\odot}$, in which range the $M_{500} - z$ space is rather uniformly covered, without
strong $z-$dependent biases. In contrast, clusters 
with $M_{500} < 10^{14}$ $M_{\odot}$ are mostly found in low-redshift subsamples, probably due to an obvious selection
effect. To quantify the effect, we projected the $M_{500}-z$ plot onto the $M_{500}$ axis to derive a mass function for each subsample. As shown in
Figure 2d, the three subsamples define similar mass functions at $>10^{14}$ $M_{\odot}$, while the subsample L contains twice more 
objects with $M_{500} \leq 10^{14}$ $M_{\odot}$ than the other two, thus reconfirming Figure 2c. To avoid this selection bias, we limit the subsequent
study to the clusters with $M_{500}$ in the range of $(0.1 - 2) \times 10^{15}$ $M_{\odot}$. This defines our ``Final Sample'' containing 340 clusters,
which are subdivided into 119, 117, and 104 objects in the subsamples of L, M, and H, respectively.

\subsection{X-ray Data}

Our Final Sample includes 141 and 199 clusters with X-ray data from {\it Chandra} and {\it XMM-Newton} archives, respectively.  

\subsubsection{\it Chandra}

The {\it Chandra} data obtained with its advanced CCD imaging spectrometer 
(ACIS) were screened with the CIAO v4.6 software and CALDB v4.6.3. Following
the standard procedure described in, e.g., Gu et al. (2009), we discarded bad pixels
and columns, as well as events with \asca\ grades 1, 5, and 7, and corrected the
event files for the gain, charge transfer inefficiency, astrometry and cosmic ray 
afterglow. By examining the lightcurves extracted in $0.3-10.0$ keV from source-free
regions, we identified and excluded time intervals contaminated by flare-like 
particle background with count rate more than $20$\% higher than the mean quiescent
value. The ACIS-S1 data were also used to crosscheck the flare detection. This usually 
reduced the exposure time by $1-3$ ks; in a few cases the flares occupy $\sim 10$ ks. 
All point sources detected above the 3$\sigma$ threshold with the CIAO tools {\tt celldetect}
and {\tt wavdetect} were masked out. The spectral ancillary response files and 
redistribution matrix files were created with {\tt mkwarf} and {\tt mkacisrmf}, respectively.

To determine the background for each observation, we extracted a spectrum from a 
region $\geq 1-2$ Mpc away from the cluster X-ray peak, and fit it with a model
consisting of an absorbed thermal component for the ICM, an absorbed power law component 
(photon index set to 1.4) for the cosmic X-ray background (hereafter CXB), and another
absorbed low temperature thermal component (temperature = 0.2 keV and abundance = 1 $Z\odot$)
for the Galactic foreground. The quiescent particle background was calculated based on 
the stowed ACIS observations (e.g., Markevitch et al. 2013). The normalizations of the 
ICM, CXB, and Galactic components, as well as the ICM temperature and abundance, were 
set free in the fitting. The best-fit sample-average CXB flux is $6.8 \times 10^{-8}$
ergs cm$^{-2}$ s$^{-1}$ sr$^{-1}$ in $2.0-10.0$ keV, which agrees well with the {\it ASCA}
result reported in Kushino et al. (2002). The combined uncertainties of the CXB and Galactic 
components were measured to be $\sim 15$\% and $\sim 30$\% for typical high-count and low-count
data, respectively. These values will be used in \S 3.2 to determine the uncertainties of 
the ICM mass profiles.

\subsubsection{\it XMM-Newton}

The SAS software v13.5.0 was used to screen and calibrate the data obtained with
the {\it XMM-Newton} European Photon Imaging Camera (EPIC). Following the 
procedure in e.g., Gu et al. (2012), we selected {\it FLAG} = 0 events, and
set the {\it PATTERN} ranges to be $0-12$ and $0-4$ for the MOS and pn data, respectively. 
Then we defined a source-free region for each observation, and extracted lightcurves
in both $10.0-14.0$ keV and $1.0-5.0$ keV bands. By investigating the lightcurves,
we discarded time intervals contaminated by flare-like background, in which the 
count rate exceeds 2$\sigma$ above the quiescent mean value in either of the two 
bands (e.g., Katayama et al. 2004; Nevalainen et al. 2005). We detected and removed
point sources with the SAS tool {\tt edetect\_chain}, and created the ancillary 
responses and redistribution matrices with {\tt xissimarfgen} and {\tt xisrmfgen},
respectively. The background model for the {\it XMM-Newton} data was determined 
in a similar way as that for the {\it Chandra} data. The EPIC spectra extracted 
from $\geq 1-2$ Mpc away from the cluster center were fit with a model combining 
the ICM, CXB, and the Galactic foreground components. The same region in a wheel-closed dataset 
was used to calculate quiescent particle background. The MOS1, MOS2, and pn spectra were fitted
simultaneously with the same model, but leaving free the cross normalizations among the three. 
The best-fit sample-average 
CXB flux is $7.2 \times 10^{-8}$ cm$^{-2}$ s$^{-1}$ sr$^{-1}$ in $2.0-10.0$ keV,
which agrees well with the {\it Chandra} result. The typical uncertainty of the X-ray
background (CXB + Galactic), $\sim 20$\%, will be later included in the error of 
the ICM mass profiles (\S3.2).

\subsection{Optical Data}

Our prime purpose of using the SDSS-III data is to select a relatively complete set of member galaxies of each cluster, and derive their radial distribution around
the cluster center. For this purpose, we used all the available photometric data.
For each target cluster in the sample, a sky region of projected radius $R < 2.5$ Mpc was defined, and all galaxies within this region were initially selected.
For each of them, we used photometric redshift (hereafter $z_{\rm phot}$), $K$-correction,
and absolute magnitude $M_{r}$ from a table $Photoz$ which is based on the method of Csabai et al. (2007). Saturated star-like objects and those
with blending problems were removed using object flags\footnote{(flags \& 0x20) = 0 and (flags \& 0x80000) = 0 and ((flags \& 0x400000000000) = 0 or psfmagerr$_r <= 0.20$) 
and ((flags \& 0x40000) = 0)}. To exclude $z_{\rm phot}$ determinations with apparently bad photometry, we discarded objects when the $z_{\rm phot}$ errors exceed $0.08 (1 + z_{\rm phot})$. This removed 
about 20\% of galaxies. Then, following Wen et al. (2009), the member galaxies of each sample cluster were selected
with a redshift filter of $z_{\rm cl} - 0.04 (1 + z_{\rm cl}) < z_{\rm phot} < z_{\rm cl} + 0.04 (1 + z_{\rm cl})$, where $z_{\rm cl}$ is cluster redshift measured with spectroscopy.
As shown in Wen et al. (2012) as well as in Figure 3, the $z_{\rm phot}$ values of the member 
galaxies in the SDSS sample determined in this way are consistent with their spectroscopic redshifts (hereafter $z_{\rm spec}$), when available, within an average scatter 
of $\approx 0.02-0.03$.

Since the SDSS data become slightly less complete in the faint end at high redshift, it is necessary to define a redshift-dependent limiting magnitude
to keep a constant completeness in the whole redshift range. Following Wen \& Han (2015b), the limiting magnitude was set as $M_{r}^{\rm e} \leq -20.5$, where $M_{r}^{\rm e}$
is the $r$ band absolute magnitude corrected for passive evolution, $M_{r}^{\rm e} (z) = M_{r} (z) + Q z$, and $Q = 1.16$
is an evolution coefficient. To estimate the possible light lost by the magnitude cut, we examined the composite luminosity function of nearby clusters based on
the WINGS survey (Fasano et al. 2006), which is complete down to $M_{V} = -15.15$. Our limiting magnitude would filter out $\approx 37$\% of the total light; when
focusing on the non-dwarf members with $M_{V} < -19$ (Moretti et al. 2015), $\approx 17$\% of the light would be lost due to the magnitude cut. This suggests that
most of the bright galaxies retain after the filtering.

To further enhance the completeness of member galaxy selection, we incorporated the $z_{\rm spec}$ values of the candidate galaxies measured with the SDSS DR12 (Alam et al. 2015).
As shown in Figure 3a, measurements of galaxy $z_{\rm spec}$ are available for $\sim 80$\% of the sample clusters, although the completeness relative to the 
photometric measurements decreases significantly towards higher redshifts. For each cluster, we selected objects for which the $z_{\rm spec}$-specified 
recession velocity is within $\pm 2500$ km $\rm s^{-1}$ of the cluster system velocity in the cluster frame.
By further applying the limiting magnitude $M_{r}^{\rm e} \leq -20.5$, a new member galaxy catalog was then created. The results of spectroscopic and photometric 
selections were combined to form the final catalog: all galaxies selected by $z_{\rm spec}$ were automatically considered as members, while the $z_{\rm phot}$
selection was used in case $z_{\rm spec}$ was not available. The sample is roughly $z_{\rm phot}$-based, since the $z_{\rm spec}$ selection contributes only
about 20\% of the total members.

For later use in \S3 when cross-checking results to be obtained with the $z_{\rm phot}$-based sample, we have also constructed a pure $z_{\rm spec}$-based sample by selecting objects with recession velocities within
$\pm 2500$ km $\rm s^{-1}$ of the cluster velocities. The $z_{\rm spec}$ sample thus has a typical galaxy number of $\sim 100$ per cluster at $z < 0.1$, while
at higher redshifts, the number becomes $\leq 50$ per cluster for most objects. In this work we focus on clusters with relatively good completeness, i.e., 
detected number $\geq 50$. This gives a $z_{\rm spec}$ subsample with 80 clusters. Details of the $z_{\rm spec}$ subsample is discussed in \S3.1.3.

We measure the galaxy luminosity function of each cluster by binning the selected members into an interval of 0.4 in absolute magnitude 
and counting the number in each bin.
To take into account the detection limit of the SDSS dataset, we corrected the derived luminosity functions by the completeness in the observed 
$r$-band. The completeness was determined by comparing the number of objects in each magnitude bin found by the SDSS, to that from the WINGS database, 
in a sky region that has been covered by both surveys. The SDSS sample is reasonably complete, at levels of $>90$\% and $\approx 85$\%, for $M_{r} < -21.5$
and $M_{r} \sim -20.5$, respectively. After applying the correction for the detection limit, we have also corrected the luminosity function for a photometric-redshift background, 
which is described in detail in \S3.1, as well as for the effect of passive evolution of red sequence galaxies. As shown in Figure 4, the obtained luminosity
functions are nearly the same among the three redshift subsamples, and agree well with the luminosity function of a more extensive SDSS sample reported in Wen \& Han (2015b).

\subsection{Determination of $R_{500}$ and the Cluster Center}
   
To compensate for the difference of the sample clusters in their physical scales, we normalize the galaxy and ICM radial profiles to a characteristic radius $R_{500}$ (see \S2.2 for definition). For this purpose, the $R_{500}$ values of the clusters in our final sample were determined through two independent approaches. First we calculated
a total mass distribution for each cluster based on the X-ray data and hydrostatic method, and determined directly its $R_{\rm 500, X}$. Details of the mass calculation 
are in \S3.3. The second method is based on an empirical scaling relation between the optical luminosity integrated over member galaxies and the cluster radius. As 
reported in, e.g., Popesso et al. (2004), the total optical luminosity $L_{\rm op}$ can be used to estimate $R_{\rm 500, op}$ by

\begin{equation}
\log R_{\rm 500, op} = 0.44 \log ( L_{\rm op} / 10^{12} L_{\odot}) - 0.15.
\end{equation}

\noindent As shown in Figure 5, the two determinations of $R_{\rm 500}$ are consistent with each other within a typical scatter of 30\%. Similar scatter is seen in other works (e.g., Wen et al. 2012). We employ $R_{\rm 500, X}$ in the subsequent study, since it is closer to the original definition.

Next, the central point for each cluster was determined. Following Paper I, we set the cluster center as the centroid of the X-ray brightness, around which the X-ray emission becomes most circularly symmetric, because the X-ray deprojection analyses (\S3.2 and \S3.3) must be precisely centralized on the X-ray centroid. The position of the brightest cluster galaxy is not employed for this purpose, because it 
can sometimes differ from the X-ray centroid by a few tens to hundreds kpc 
(Shan et al. 2010) in cases of merging clusters.
As shown in Paper I, the X-ray centroid was calculated in an iterative way for each cluster. 
By using a CIAO task {\tt dmstat} on the central $r_{\rm 1} = 5^{\prime}$ region, the first centroid
was derived. Then we carried out iterations by reducing gradually the radius $r_{\rm i+1} = r_{\rm i} - 1^{\prime}$, and calculated the new centroid
as well as the shift from the previous one. The iteration was considered as converged when the shift became less than $10^{\prime\prime}$. The X-ray centroid
is compared in details to the optical center in \S3.5.1.

\section{DATA ANALYSIS AND RESULTS}

\subsection{Galaxy Number Density Profiles}

For each cluster, the galaxies selected with the $z_{\rm phot}$-based method (\S2.4) were combined to calculate surface number density profiles.
These profiles might still contain a residual background component due to possible false member selection with $z_{\rm phot}$. To remove such a component, 
we calculated the mean galaxy density in a surrounding region, i.e., between 4 Mpc
to 8 Mpc from the cluster center. To exclude possible large-scale structures in the background region, we further divided it into 48 sectors with equal area. Those
sectors with galaxy densities larger than 2$\sigma$ of the mean value were discarded, and a new mean background density was then obtained from
the remaining sectors. The same method was used in Popesso et al. (2004). After subtracting the background for each cluster, we 
normalized the density profile by dividing the radius $r$ by $R_{500}$. This compensates the difference of cluster scale along the sky plane; to 
further correct it in the line-of-sight direction, we further divided the surface galaxy number density by $R_{500}$.

In Figure 6, we show the sample-averaged radial number density profile of galaxies in comparison with the one reported in Budzynski et al. (2012; B12 hereafter). The B12 profile was
obtained by stacking over 50000 clusters and groups in $0.15 < z < 0.4$, and the number density of each cluster was calculated
by a direct background subtraction approach instead of member selection. Though derived with quite different methods, 
the two profiles nicely agree with each other up to 2$R_{500}$.

\subsubsection{Error modelling}

In order to estimate uncertainties in the galaxy number density profiles, we considered two error sources: the Poisson error on 
the number of selected member galaxies in each radial bin, and the systematic uncertainty on the $z_{\rm phot}$ measurements. The former 
was calculated using the formulae of Gehrels (1986), while the latter was determined using a Monte-Carlo simulation in the redshift
space. For each cluster field, $z_{\rm phot}$ of all galaxies (both members and non-members) were shifted randomly within
the measurement error, i.e., $\sigma_{z} = 0.03$ and 0.07 for objects with $z_{\rm phot} < 0.5$ and $z_{\rm phot} > 0.5$, respectively.
Then, we re-selected member galaxies by using the method in \S2.4, and re-calculated the number density profiles. By 1000 Monte-Carlo
realizations for each cluster, we determined the systematic uncertainty to be associated with the $z_{\rm phot}$ measurement. Typically the 
systematic error is larger than the Poisson one by a factor of $\sim 5$. The two
errors were combined in quadrature to make up the total error. As shown in Figure 6, the combined uncertainty is $\sim 8$\% in the 
central bin, and $\sim 10$\% at outer radii.

\subsubsection{Dependence on cluster mass and redshift}

It has been expected that the dark matter and member galaxy distributions should in general show self-similarity, i.e., the 
density profiles of different systems become nearly identical after some scaling (e.g., Navarro et al. 1997). To explore any additional processes
beyond this simple self-similarity, we examine the possible mass- and redshift- dependence of the galaxy density profiles. First
we split the sample into three mass groups, i.e., $1 \times 10^{14} - 2 \times 10^{14}$ $M_{\odot}$, $2 \times 10^{14} - 5 \times 10^{14}$ $M_{\odot}$, and
$5 \times 10^{14} - 2 \times 10^{15}$ $M_{\odot}$. Details of the cluster mass calculation are given in \S3.3. 
The choice of the three mass groups is to have similar number of clusters in each group.
The average number density profiles of the three mass groups are shown in Figure 7b. The scaling by $R_{500}$ has removed most of the mass
dependence in the galaxy distribution. The only small difference is seen at the second radius bin, where the high mass profile
is $\sim 20$\% lower than the other two profiles.

The essence of Paper I was a discovery of significant evolution of the member galaxy distribution, relative to that of the ICM. As a major step to 
confirm this discovery, we calculated 
number density profiles individually for the entire sample clusters, and then took their ensemble averages over the redshift-sorted subsamples L, M, and H (\S 2.1). 
Unlike the case of mass-sorting, the three galaxy distributions reveal, as shown in Figure 7d, a clear evolution; 
towards lower redshifts, the profile becomes systematically steeper, with a stronger central increment within $0.2 R_{500}$, 
and a quicker outward drop at $> 0.5 R_{500}$. The average 
ratio between the subsamples L and H is $\sim 1.6$ at the central bin, and $\sim 0.8$ at the outermost.  
A similar result was reported in Ellingson et al. (2001). As shown in their Figure 9, the populations of red-sequence and field-like + post-star formation galaxies both become more concentrated towards lower redshifts.

\subsubsection{$z_{\rm spec}$ sample}

To crosscheck the above results, we repeated the same analysis on
the $z_{\rm spec}$-based sample constructed in \S2.4. Compared to the $z_{\rm phot}$-based sample, the spectroscopic one 
gives more reliable determination of cluster members, but it is more biased to high-flux galaxies ($r < 19.2$; e.g., Eisenstein et al. 2001).  
As described in \S2.4, the $z_{\rm spec}$ sample was constructed to include 80 clusters. Then, for each cluster, we calculated in the same way the radial
profile of surface galaxy number density. As shown in Figure 8, the galaxy profiles are averaged over
two redshift ranges, $z < 0.075$ and $z>0.075$. Error bars show only the Poisson error on the number of 
galaxies. The two averaged profiles clearly exhibit different gradients: 
the nearby one is more centrally-peaked than the distant counterpart. The central bin of nearby clusters
is on average $\sim 40$\% higher, while the outer bins are $\sim 20$\% lower, than those of distant objects. 
Compared to the $z_{\rm phot}$-measured profiles (Fig. 7d), the $z_{\rm spec}$ ones are systematically lower by
$\sim 20 - 50$\% due to the relatively low completeness. The redshift-resolved gradients of the galaxy profiles 
are found consistent between the two $z_{\rm phot}$ and $z_{\rm spec}$ samples.

\subsubsection{Concentration}

To quantify the difference in the member galaxy distributions, we measured the concentrations of the density
profiles as described by a standard NFW model (Navarro et al. 1997), which can be written as
\begin{equation}
\rho (r) = \frac{\rho_0}{\frac{r}{R_{\rm s}}\left(1+\frac{r}{R_{\rm s}}\right)^2}.
\end{equation} 
Here $\rho (r)$ is the galaxy  density, $\rho_0$ is the normalization, and $R_{\rm s}$ is the scale radius.
Since the galaxy profiles typically extend out of $R_{500}$, the ratio $c_{500} = R_{500} / R_{\rm s}$ is used here
as the concentration parameter. The same or similar definition is often used to describe the structure of dark matter halos (e.g., 
Vikhlinin et al. 2006; Dolag et al. 2004). Then we projected the NFW model along line-of-sight, and fit the galaxy 
profiles in Figure 7d over the radius of $0.05 R_{500}$ to $2.4 R_{500}$. The best-fit concentrations are $2.5 \pm 0.2$,
$2.1 \pm 0.1$, and $1.8 \pm 0.1$ for the subsamples L, M, and H, respectively, re-confirming the evolution
found in \S3.1.2. The range of the concentration parameter agrees with those reported in previous works (e.g., 
Lin et al. 2004; Muzzin et al. 2007; B12)

\subsection{ICM Density Profiles}
Next we address whether or not the same evolution is present in the ICM component.
To determine the ICM mass distribution, the 3-D ICM density profile was calculated for each cluster in a standard way based on deprojection spectral analysis with 
the {\it XMM-Newton} and {\it Chandra} data (e.g., Paper I). After removing point sources, we extracted spectra from several concentric annulus regions
for each cluster. The radial boundaries of each annulus were determined to include sufficient net counts, i.e., 1000 and 3000 for 
low-flux and high-flux clusters, respectively. The extracted spectra were fit in $0.7-8.0$ keV with an absorbed, single-temperature APEC
model in {\tt XSPEC}. All annuli were linked by a PROJCT model, which assumes the plasma parameters (e.g., emission measure and temperature) of the
corresponding 3-D shells individually, calculates via projection a set of X-ray spectra to be observed from the specified 2-D annuli, and compares them with the actual data. For each 3-D shell, 
the gas temperature and metal abundance were set free. When the model parameters cannot be well constrained in
some annuli due to relatively low statistics, we tied them to the values of their adjacent regions. The column density 
of neutral absorber was fixed to the Galactic value given in Kalberla et al. (2005). We have also 
calculated spectral parameters with a more direct deprojection method presented in Sanders \& Fabian (2007), which gave consistent results with the PROJCT
method. All the fits were acceptable,
with reduced chi-squares $\sim 0.8-1.2$ for a typical number of degrees of freedom being $\sim 300-1000$.

The 3-D ICM density profile was then calculated
from the best-fit model normalization of each annulus. Errors was estimated by taking into account both statistical and systematic 
uncertainties. The former was estimated by scanning over the parameter space with an {\tt XSPEC} tool {\it steppar} iteratively, while the latter
was assessed by renormalizing the level of CXB component for each region by 20\% (\S 2.3). The two kinds of errors are comparable for most
annuli. The sample-average gas density profile is then obtained and plotted in Figure 9. It decreases from $\sim 10^{-2}$ cm$^{-3}$ in the central 
100 kpc, to a few $10^{-4}$ cm$^{-3}$ at $r > 1$ Mpc. In the same plot, our profile is compared with a
previous result by Croston et al. (2008), which reports the ICM density profiles of 31 nearby clusters ($z<0.2$). The
two results agree nicely with each other at all radii.

As shown in Figures 10a and 10b, the $R_{500}$-scaled ICM density profiles were firstly grouped by the cluster mass as introduced in \S3.1.2. At small to intermediate radii 
($<0.5 R_{500}$), vertical separation can be 
clearly seen among the average profiles of the three mass groups: the value of the high-mass group at $0.3 R_{500}$ is larger than those of the medium-mass and low-mass ones by a factor of $\sim 1.4$ and 1.6, respectively. At larger radii, however, the dependence becomes weaker, and the three profiles are 
consistent at $\geq R_{500}$. Such a feature agrees well with those reported in previous observations (e.g., Croston et al. 2008) and 
simulation works (e.g., Borgani et al. 2004).

Then we investigate how the spatial distribution of hot ICM evolves with time. As shown in Figures 10c \& 10d, the scaled ICM density profiles were binned, as before, by their 
redshifts. In the inner region, the average profiles of the three subsamples show weak dependence on redshift, in contrast to the cluster mass-sorting. The core density at $0.1 R_{500}$ 
decreases by $\sim 10$\% from subsample H to subsample M, and by $\sim 15$\% from subsample M to subsample L. Similar to the case of the mass-sorting,  
this correlation becomes further weaker and vanishes at $>0.5 R_{500}$. The evolutionary trend of ICM density profiles is opposite to that found in Figure 7d in the galaxy density profiles.

\subsection{Total Gravitating Mass}

To study the most dominant mass component, the dark matter, we calculated the total gravitating mass profile for each cluster. Here we employed 
the standard hydrostatic mass estimates using the X-ray data (e.g., Sarazin 1988; M01). Based on the best-fit 3-D gas temperature profiles $T_{\rm
  X}(R)$ and density profiles $n_{\rm g}(R)$ obtained with the
deprojected analysis (\S 3.2), and assuming spherical symmetry and a
hydrostatic equilibrium, the total gravitating mass within a 3-D radius $R$
can be calculated generally as
\begin{equation}
M(R) = \frac{-kT_{\rm X}(R) R}{G \mu m_{\rm p}}\left( \frac{d {\rm ln}
  n_{\rm g}(R)}{d {\rm ln} R} + \frac{d {\rm ln} T_{\rm X}(R) }{d {\rm
    ln} R}\right),
\end{equation}
where $G$ is the gravitational constant, $\mu = 0.609$ is the assumed average
molecular weight, and $m_{\rm p}$ is the proton mass. The associated errors were calculated combining 
those on the temperature and the density. By stacking the mass profiles obtained from individual clusters, we 
show the sample-average mass profile in Figure 11. 
The radius $R$ and mass $M(R)$ are normalized to $R_{500}$ and $M_{500}$, respectively.

To verify our mass measurements, we compare the results derived in this way with those reported in previous X-ray and weak lensing 
studies. As shown in Figure 11, our sample-average mass profile agrees nicely with those presented in Zhang et al. (2008),
which were calculated with the same X-ray technique for 37 clusters at $z = 0.14-0.3$. Most objects in their sample are also present 
in this work. The profiles of the two samples, this work and Zhang et al. (2008), also 
give similar scatter, $\approx 0.25$ and $\approx 0.2$ of the mean value at $r = 0.3 R_{500}$. For a further test, we compared our 
mass profiles with those obtained in previous weak lensing studies (e.g., Dahle et al. 2002). Since the lensing gives
aperture mass projected along the line-of-sight, we have projected the X-ray mass profiles which were originally obtained in 3-D.  
The X-ray and weak lensing results are generally consistent with each other in most radii, albeit with 
some differences seen at central $\sim 0.1 R_{500}$ in several objects. These discrepancies cannot affect the key profiles 
shown in \S3.4, which have innermost bins at $\sim 0.25 R_{500}$.

Having derived the X-ray mass profiles, let us investigate how the matter distribution varies with the cluster mass $M_{500}$. The mass profiles for clusters in the three mass groups (\S 3.1.2) are plotted in Figures 12a and 12b in different colors. The total gravitating masses of the low-mass clusters thus exhibit higher concentration than those of the high-mass
ones. To quantify this effect, we again calculated the concentration parameter as $c_{500} = R_{500} / R_{\rm s}$ (\S 3.1.4), where $R_{\rm s}$ is the scale radius 
determined by fitting the mass density profiles with the NFW model (Eq.2). The central region $r < 0.05 R_{500}$ was excluded in the fitting, because the mass density at these radii are often found biased from the universal model (e.g., Gu et al. 2012). The obtained $c_{500}$ is plotted as a function of $M_{500}$ in Figure 13. Thus, $c_{500}$ decreases significantly as a function of mass; the average value of high-mass clusters is about two-thirds of that of low-mass ones. Our result is generally consistent with the expected $c - M$ relation by cluster simulations (Dolag et al. 2004) and those from previous X-ray observations (e.g., Vikhlinin et al. 2006).  

Following this, the total gravitating mass profiles were regrouped in redshift. As shown in Figure 12d, the average profiles of the three subsamples in general trace each other,
and shape of the profiles does not vary strongly across the redshift range. The mean value of the subsample L is shifted to be slightly lower, by $20-30$\%, than those of the distant subsamples, probably due to the sample selection effect as noticed in Figure 2d. Combining Figure 12b and Figure 12d, it is suggested that the concentration of dark matter halo depends primarily on the
cluster mass rather than the redshift.

\subsection{Comparison of Galaxy/ICM/Dark Matter Distributions}

To compare directly the spatial distributions of the three mass components, here we calculated three types of radial profiles, 
i.e., galaxy number vs. ICM mass ratio (hereafter GNIMR), galaxy number vs. total mass ratio (hereafter GNTMR), and ICM mass vs. total
mass ratio (hereafter IMTMR). As a first step, the obtained radial profiles of the three mass components were transformed
to have the same form and dimension. For each cluster, the galaxy number density enclosed in each 2-D radius $r$ (\S 3.1) were integrated to obtain
a quasi-continuous integral profile. To match with the optical profile, the ICM mass profile was calculated by projecting numerically the ICM density
distribution (\S 3.2) along the line-of-sight, and integrating over the same set of radius bins. As for the total mass, we converted the mass profile (\S 3.3) into the 3-D mass density profile, 
$\rho (R) = (4 \pi R^{2})^{-1} dM(R) / dR$, projected $\rho(R)$ onto the sky plane, and integrated it over each radius $r$ to
obtain a 2-D mass profile. The radially-integrated profiles of the three components, now in the same form and dimension, were further normalized to their 
central value at $r=0.25 R_{500}$ ($\sim 250$ kpc), because we are interested in their relative shape differences. This radius was chosen to be in line with paper I. Finally, by dividing one component to
another, we obtained the GNIMR, GNTMR, and IMTMR profiles for each cluster. The ratio profiles were again averaged over subsamples
as a function of either the cluster redshifts or the masses, and are shown in Figure 14. Uncertainties of the averaged profiles were
calculated by combining in quadrature the errors of all clusters in the subsample.

\subsubsection{Dependence on cluster mass and redshift}

As shown in Figure 14b, the GNIMR profiles exhibit apparent evolution with the redshift: the average profile of subsample H is significantly
flatter than that of subsample L, and that of subsample M shows an intermediate gradient. This result is a direct consequence of the opposite evolutionary 
trends revealed in Figure 7d and Figure 10d. Furthermore, it is consistent with the one
reported in Paper I (in its Fig. 13b), which is obtained with a very different galaxy selection method. As expected, the high-redshift ($z=0.4-0.9$)
profile of Paper I appears to be clearly flatter than that of subsample H of this work. This indicates that the observed evolution might extend 
continuously to high redshifts. Furthermore, the shape of the GNIMR profile does not 
depend strongly on the cluster mass (Figure 14a), which proves that the redshift dependence cannot be attributed to the obvious selection bias
that we tend to select more massive objects at higher redshifts (\S 2.2).

Similar to the GNIMR, the shape of the GNTMR profile also depends clearly on redshift, as shown in Figure 14d. While the galaxies used to be less concentrated
than the underlying dark matter in subsample H, they have evolved to become slightly more concentrated than the latter in subsample L. As can be 
inferred by comparing Figure 7d, Figure 10d, and Figure 12d, the redshift dependences of the GNIMR and GNTMR profiles are mainly contributed by 
the z-dependent changes in the galaxy number 
distributions. On the other hand, as shown in Figure 14c, the GNTMR profiles do not depend significantly on the cluster mass in the central $0.7 R_{500}$. The only difference is seen in the outer radii, where the low-mass clusters show relatively flatter GNTMR than the high-mass ones. This is mainly
due to the mass-concentration effect of the dark matter halos (Fig. 13).

The ICM exhibits the most extended distribution among the three components. Furthermore, as shown in Figures 14e and 14f, the IMTMR profile depends on both 
redshift and the cluster mass: the X-ray size relative to that of dark matter increases with decreasing redshift and cluster mass. While
the two dependences are both caused mainly by the behavior of the ICM density profiles shown in Figures 10b and 10d, the mass dependence is also 
contributed by the trend of mass concentration (Fig. 13).

Figure 15 provides another presentation of the above results, where the subsamples (either redshift- or mass- sorted) are plotted  
on the GNTMR vs. IMTMR plane. Thus, relative to the dark matter, the stellar component 
kept shrinking from $z=0.5$ to $z=0$, while the ICM component evolved to be more and more extended even though it suffers strong radiation
loss. Any mass-related effect cannot be a major driving force of this evolution, since the mass-sorted subsamples are seen to
behave in quite different way on the same plot.

\subsubsection{Galaxy light vs. ICM mass ratio}

In order to examine consistency with Paper I, we may need to slightly modify the GNIMR calculation by replacing the galaxy number profile with a galaxy light profile.
We calculated the rest-frame $r$-band luminosity of all galaxies selected by $z_{\rm phot}$, corrected it for
passive evolution (\S2.4), subtracted a residual background to eliminate possible false selection (\S3.1), and integrated the galaxy light in each
radius. The galaxy light vs. ICM mass ratio (hereafter GLIMR) profile was then obtained for each cluster. As shown in Figure 16, 
the subsample-averaged GLIMR profiles exhibit strong evolution in gradient as a function of redshift; the distribution of stellar-to-ICM ratio is more 
concentrated in lower redshifts. This feature is in good agreement with those obtained in paper I. When compared to Figure 14b, the GLIMR profiles appear to be 
steeper than the GNIMR ones in subsample L and M, indicating that the brighter galaxies, hence more massive objects, preferably reside in the central regions for
the nearby clusters. This feature is less significant in subsample H.

\subsection{Systematic Errors and Biases}

Here we examine in details the possible systematic errors and biases that might be involved in the galaxy-to-ICM comparison.

\subsubsection{Selection of cluster center}

In the above studies, the center of each cluster was defined as the X-ray centroid (\S2.5). It is however known that the X-ray center could
be offset from the true bottom of the potential, if, e.g., the ICM is not in hydrostatic equilibrium. An obvious alternative is to 
identify the cluster center with the position of the brightest cluster galaxy (although this could also be biased). To assess the possible bias due to 
the center selection, 
we have calculated another set of galaxy number profiles, and hence the GNIMR profiles, by re-setting the center to the brightest 
cluster galaxy of each cluster. For the reason described in \S2.5, the ICM mass profiles, centralized on the X-ray centroid, were kept unchanged. 
As shown in Figure 17a, the new GNIMR profiles appear to be systematically steeper than the original ones. This is because
the new galaxy number profiles, with optically-defined center points, often give a higher value in the inner regions. Nevertheless, the main 
result obtained in \S 3.4.1, i.e., the redshift-dependence
on the gradients of the GNIMR profiles, remains intact after the center shifting.

\subsubsection{X-ray background}

The largest source of systematic error on the X-ray profiles is uncertainties in the background subtraction, and the 
effect must be severer in X-ray faint clusters. If the X-ray background was systematically over-subtracted in low-flux
clusters, the ICM density at large radii would be suppressed, and the GNIMR profiles would thus be flattened. 
To examine this possible effect, we calculated the X-ray source-to-background ratio at 0.6 $R_{500}$ ($S/B_{0.6}$ hereafter) 
for each cluster. The mean ratios are 2.2, 2.2, and 2.1 for subsample L, M, and H, respectively, which show no redshift dependence. 
These values are larger than the limiting $S/B$ in previous studies (e.g., $S/B^{\ast} = 0.6$, Leccardi \& Molendi 2008).
To further clarify the background effect, we divided each subsample into two groups by the mean source-to-background ratio,
$S/B_{0.6} = 2.2$. As shown in Figure 17b, such effects of the X-ray background is actually visible to some extent, but the GNIMR profiles of 
the low $S/B$ and high $S/B$ groups are still consistent with each other in each subsample, 
and the evolution discovered in \S 3.4.1 is seen in both groups. This indicates that the observed evolution cannot be explained by 
background uncertainties of the X-ray data.

\subsubsection{Cosmological growth}

As noticed in paper I, there is one additional effect involved in the cluster evolutionary study.
Since clusters are considered to grow via matter accretion onto their outskirts, the cluster
scale (i.e., the mass and radius) should increase continuously over cosmological timescales. This makes
$R_{500}$ of each cluster a time-dependent quantity. To compensate for such underlying differences in the cluster evolution 
stage, we have attempted to re-calculate the cluster profiles by defining a new scale, $R_{500}^{\rm z=0}$, 
the expected $R_{500}$ that a cluster will achieve after evolving to $z=0$. Based on the empirical cosmic growth function
derived from N-body numerical simulation, e.g., Wechsler et al. (2002), the cluster mass is expected
to increase by a factor of $\sim e^{1.33 z_{0}}$ from $z = z_{0}$ to $z = 0$, so that $R_{500}^{\rm z=0}$ 
can be estimated by $R_{500}^{\rm z=0} \approx R_{500}^{z=z_0} \times e^{0.56z_0} E(z_0)^{0.58}$, where $E(z_0)$
is the cosmological evolution factor. The new radial scale is utilized to calculate  
a new set of GNIMR profiles shown in Figure 17c. Although the profiles, especially those of the high redshift subsamples,
are modified to be slightly steeper,
the new result is still consistent with the original one within error bars, and the GNIMR evolution still remains significant. The systematic uncertainties 
caused by such a ``cosmological growth'' effect in cluster outer regions should be $\leq 10$\%.

\subsubsection{Dynamical states}

Since the sample clusters have a wide scatter in morphology, it is important to examine whether or
not the detected evolution on galaxy-to-ICM profiles is created by the evolving dynamical state of clusters. 
Based on the spatial distributions of member galaxies, we can divide the sample into merging and relaxed clusters,
while the X-ray data allow us to define cool-core and non-cool-core objects. Both of these classifications can be
considered to represent the dynamical state of clusters (e.g., West et al. 1988; Allen et al. 2008). First we
utilized the two-dimensional galaxy distributions. As shown in Wen \& Han (2013),
unrelaxed clusters tend to exhibit asymmetrical galaxy distribution and bumpy brightness profiles in outer regions, while
 relaxed ones are on the opposite. Such features are quantified by three parameters, i.e., asymmetry factor $\alpha$,
ridge flatness $\beta$, and normalized deviation $\delta$ defined in Eqs.(7), (9), and (12) of Wen \& Han (2013),
respectively. As described in Eq.14 of their paper, the three factors can be further combined into one characteristic 
parameter $\Gamma$, which is most sensitive to the dynamical state. By applying the same analysis, we define a relaxation 
type for each cluster. As seen in Figure 18, this provides a clear
separation of the sample into 46\% relaxed and 54\% unrelaxed clusters. The GNIMR profiles were
rebinned by the dynamical type for each redshift-dependent subsample. As shown in Figure 17d, the differences
between the two types are $\leq 10$\%, $\leq 20$\%, and $\leq 8$\% for the subsample L, M, and H, respectively. 
Although subsample M is somewhat affected, the evolution suggested in \S 3.4.1 is still clearly seen in both types.

Then we crosscheck the above result with the X-ray data. 
It is well known that the relaxed clusters frequently possess cool ICM cores, while merging clusters
have relatively flat cores (e.g., Santos et al. 2010). Hence, the sample should be separated into cool-core and non-cool-core
subsamples. Following Sanderson et al. (2006), we measured the core temperatures
using spectra of the $<0.1 R_{500}$ regions, and the cluster mean temperatures in the $0.1-0.2 R_{500}$ regions.
We defined cool-core objects as those systems in which the mean temperature exceeds the core temperature at $>3\sigma$
significance. The sample was thus divided in to 44\% cool-core and 56\% non-cool-core objects. As shown in 
Figure 18, this X-ray classification is in fact close to the optical method.
This is probably because that the optical $\Gamma$ factor has been somewhat calibrated to the X-ray sample 
(Wen \& Han 2013). Then, we rebinned the GNIMR profiles by the ICM core state (Fig. 17e). Again, the GNIMR profiles
are not strongly affected by this operation, and the resulting variations are $\leq 6$\%, $\leq 12$\%, and $\leq 5$\%
for the subsample L, M, and H, respectively. The detected evolution remains intact for both cool-core and non-cool-core
objects.

\subsubsection{Galaxy color}

Another major concern with the GNIMR evolution is possible redshift-dependence of member galaxy type (e.g., color), 
coupled with the decreasing detecting efficiency of blue galaxies towards higher redshift (e.g., Csabai et al. 2003). 
To assess the effects of galaxy evolution, we here divided our sample, by a color-magnitude diagram employing the $(g-r)$ color, 
into red sequence galaxies and blue galaxies. 
To determine the color-magnitude relation for each cluster, first we calculated the zeropoint
of the $(g-r)$ vs. $r$ diagram by fitting it with a straight line, assuming a uniform slope of -0.02 for all redshifts.
This is valid because the slope is primarily a result of metallicity and has little evolution
with the stellar age (e.g., Kodama et al. 1998).  We utilized a
robust biweight linear least square method (Beers, Flynn \& Gebhardt 1990), which is insensitive
to data points that are much deviated from the relation. The fitting was repeated for a few times, as 
the data points outside the 3$\sigma$ range of the fitted relation were rejected in the next
iteration. In performing the fitting, we included those cluster members which have been confirmed by spectroscopic measurements. 
The best-fit zeropoint increases from 1.1 to 2.5 towards high redshift. The
zeropoint vs. redshift relation is in good agreement with the metallicity sequence model reported in Kodama \& Arimoto (1997).

Then, we created color-magnitude diagram of all member galaxies of the $z_{\rm phot}$ sample, and selected
red-sequence members as those within $\pm 0.15$ magnitudes of the best-fit color-magnitude relation.
This simultaneously defines the blue members as those in the blue cloud. Figure 19 shows six examples of 
the color-magnitude selection. Since the color cut extends to the blue side below the color-magnitude relation, most red-sequence
galaxies have been selected, although the red sample is inevitably contaminated by blue galaxies in
the faint end. To examine the effect of blue contamination, we carried out two independent approaches. First, we 
selected galaxies only on the redder side of the best-fit color-magnitude relation, and mirrored the distribution 
about the relation to determine the blue side. A similar
selection was used in Gilbank et al. (2008). This affected the average galaxy surface number density profiles only
by $\sim 5$\%. Second, we excluded the faint end, $M_r > -20$, from the red galaxy sample. Although
the resulting number density profiles are systematically lower in the second approach, the slopes of the profiles 
are consistent within $1\sigma$ between the two methods. Hence, we applied the first approach to separate the red and blue galaxies.

Next we examine the possible evolution in the spatial distributions of red and blue galaxies. In Figure 20
we show the surface number density profiles of red and blue galaxies as a function of redshift. The red-sequence members have a higher density,
together with a more centrally-peaked distribution, than the blue galaxies detected within $r_{500}$. This feature has been reported
in many previous works, e.g., Kodama et al. (2005), Koyama et al. (2011). When comparing the three redshift bins, the red members exhibit clear evidence of 
evolution which is analogous  
to that of the total profiles (Fig. 7d); the nearby clusters show more peaked galaxy distributions than the distant ones. 
As for the blue component, the similar feature is again seen in the cluster outer regions ($r>0.2 R_{500}$). Towards the center, 
the blue 
profiles of the three redshift subsamples do not show marked differences, within rather large errors which are probably due to the
relatively poor capacity of the blue galaxy selection with $z_{\rm phot}$. As shown in Figure 17f and 17g, similar properties are present on the GNIMR profiles of
the red and blue galaxies; both exhibit signs of evolution, while those of the blue galaxies are overpowered 
by error bars. Hence, we prove that the observed redshift-dependence on the galaxy-to-ICM profiles cannot be fully ascribed to 
the time evolution of the galaxy colors.

\section{DISCUSSION}

\subsection{Evolution of the GNIMR/GNTMR/IMTMR Profiles}
  
By analyzing the optical and X-ray data of a large, X-ray bright cluster sample with $z<0.5$, we studied 
radial distributions of the three major mass components, i.e., the member galaxies, the hot ICM, and dark matter.
The galaxy membership was determined primarily with the photometric redshift. The total 340 clusters were grouped
into three subsamples (L, M, and H) by their redshifts. The GNIMR and GNTMR profiles were found to drop towards outer regions, with a 
slope that clearly steepens from subsamples H through M to L. The IMTMR profiles exhibited opposite (positive) slopes
and an opposite evolution compared to GNIMR. The behavior of GNTMR was explained as a composite of the above two effects. 
The combined evolution on GNIMR, GNTMR, and IMTMR profiles cannot
be mimicked by mass-related effects (Figs. 14 and 15). As shown in Figure 16, these results quantitatively confirm and substantially expand the discoveries
of Paper I, which is based on 34 galaxy clusters with $z = 0.1-0.9$. 

In \S 3.5, we examined all conceivable sources of systematic errors and biases that could potentially produce, 
as artifacts, the apparent GNIMR/GNTMR/IMTMR evolution. However, none of them was found to affect 
the observed GNIMR profiles by $>20$\%; the detected evolution remains intact. All the pieces of evidence consistently 
indicate that the radial distribution of member galaxies has indeed been evolving to be more centrally-concentrated relative
to the ICM and dark matter components, while the ICM has slightly expanded relative to dark matter in spite of its strong radiation loss.

The behavior of GNIMR/GNTMR could be grossly explained in at least three different ways. First, the higher galaxy density 
in the centers of nearby clusters could be directly created by enhanced star/galaxy formation therein towards lower redshifts. 
However, this idea is opposite to the currently understood evolution in the star forming rate, which is thought to have decreased 
from $z \sim 2$ (e.g., Tresse et al. 2007). Furthermore, this scenario cannot explain the central decrease of the iron-mass to 
light ratio (IMLR; the mass of iron in the ICM to the galaxy light), observed in some nearby galaxy groups (Kawaharada et al. 2009).
Second, the GNIMR evolution could be explained by successive addition of primordial gas onto the outermost periphery of 
individual clusters as they grow up. However, this disagrees with the recent X-ray results that the ICM of nearby clusters 
is metal-enriched uniformly out to their virial radii (Fujita et al. 2008; Werner et al. 2013) with a rather constant IMLR 
(Sato et al. 2012). A third idea is hierarchical cluster growth combined with static galaxy evolution. As clusters grow up from inside out, 
their outer regions would contain more newly-formed galaxies, which should be optically dimmer due to the decreasing star formation rate 
towards lower redshifts. Then, nearby clusters would have lower galaxy densities in the outer regions (relative to dark matter) than distant objects.  
However, this view can explain neither the observed gradual increase with time in the galaxy density at the cluster central regions (Fig. 7d),
nor the evolution of the number density profiles of old population galaxies (i.e., red sequence, Fig. 20).

We are hence left with the fourth view, first proposed by Makishima et al. (2001) and reinforced in Paper I, 
that the member galaxies have actually been falling, relative to the ICM and dark matter, towards the cluster 
center on a cosmological timescale. At the same time, the ICM is considered to have somewhat expanded relative to 
the dark-matter distribution. This simple yet so-far unexplored dynamical scenario can explain all essential results presented in \S 3.
Furthermore, it can explain some important features of the metal distribution in the ICM, namely, the central decrease and outward 
flatness of IMLR. However, the postulated infall of galaxies clearly requires dissipation of their dynamical energies.
In the next two subsections, we consider two possible origins of such energy dissipation.

\subsection{Dynamical Friction}

As a mechanism which causes galaxies to fall to the cluster center, we first consider dynamical friction,
which occurs as gravitationally-induced energy exchange between a moving galaxy and surrounding cluster media including dark matter,
ICM, and other galaxies/stars (Dokuchaev 1964; Rephaeli \& Salpeter 1980; Miller 1986). As shown in, e.g., El-Zant et al. (2004),
the dynamical friction can create a strong energy flow of $10^{44}$ erg $\rm s^{-1}$ per cluster out of the member galaxies,
and effectively drag the galaxies inward. To be quantitative, consider a model galaxy with a mass $m$, on a circular orbit 
with a radius of $R$ and an orbit velocity of $v=\sqrt{{\rm G} M(R)/R}$, where $M(R)$ is the gravitating mass of the cluster inside $R$.
Due to the gravitational interaction, the galaxy receives drag force as
\begin{equation}
F_{\rm DF}(R) \approx \frac{4 \pi \rho(R) ({\rm G} m)^2 }{v^2}
\end{equation}  
(e.g., Ostriker 1999; Nath 2008), where $\rho(R)$ is the cluster mass density. The angular momentum of the galaxy, $L \sim m v R$, 
decreases with time by ${\rm d}L/{\rm d}t \sim F_{\rm DF}(R) \times R$, so that the galaxy moves in a spiral trajectory with 
the radius changing by 
\begin{equation}
\frac{{\rm d}R}{{\rm d}t} \approx \frac{4 \pi \rho(R) {\rm G}^2 m R}{(k+1)v^3}, 
\end{equation}
where $k \equiv {\rm d} {\rm ln} v / {\rm d} {\rm ln} R$ is the logarithmic velocity gradient.  
Employing typical parameters, the infall distance can be written as 
\begin{equation}
\Delta R \approx 10    \left( \frac{\rho}{10^{-4} M_{\rm \odot}{\rm pc}^{-3} } \right)    \left( \frac{m}{10^{12} M_{\rm \odot}} \right)   \left( \frac{R}{500{\rm kpc}} \right)     \left( \frac{\Delta t}{10^{9} {\rm yr}} \right)   \left(\frac{v}{1000 {\rm km/s}}\right)^{-3}  {\rm kpc}.
\end{equation}
As the most prominent characteristic of this mechanism, more massive galaxies are thus predicted to fall faster
than less massive ones.

To investigate the effect of dynamical friction, it is hence important to examine whether or not the GNIMR profiles 
depend on the galaxy mass. By adopting the observed total mass-to-light vs. luminosity relation given
in Cappellari et al. (2006; Eq.9 therein), we calculated the mass $m$ for each galaxy. Then, all galaxies were divided 
into two groups by a limiting mass $m^{\ast} = 1 \times 10^{11}$ $M_{\rm \odot}$, below which the effect of dynamical
friction becomes negligible (with the infall rate $\sim 2$ kpc per Gyr; Eq.6). Figure 21 shows the GNIMR 
profiles as a function of redshift for the less massive member galaxies. Although the significance is slightly lower 
than the original result shown in Figure 14b, the average GNIMR profiles of the three subsamples are still clearly separated 
from each other. Therefore, the dynamical friction between individual galaxies and the host cluster cannot be the sole mechanism to explain 
the observed galaxy-to-ICM distribution. This reconfirms a conclusion already derived in Paper I.

\subsection{ICM Effects}

Besides the gravitational effects considered in \S4.2, galaxies are also affected by their direct interaction with the ambient ICM. One
well-known ICM effect is ram pressure caused by motion of galaxies through the ICM (Gunn \& Gott 1972). The ram pressure
force on a single galaxy is written as
\begin{equation}
F_{\rm RP}(R) \approx \pi R_{\rm int}^2 \rho_{\rm ICM}(R) v^2 
\end{equation}
(Sarazin 1988), where $R_{\rm int}$ is the effective interaction radius of the moving galaxy, and $\rho_{\rm ICM}(R)$ is the ICM density distribution. Since the galaxies and ICM have similar specific energy per unit 
mass, while galaxies have much lower specific entropy, the free energy would flow from galaxies to the ICM. 
As a result of the continuous ram pressure, the galaxy orbit will decay by
\begin{equation}
\Delta R \approx 10    \left( \frac{R_{\rm int}}{5{\rm kpc}} \right)^{2}     \left( \frac{n_{\rm ICM}}{10^{-3} {\rm cm^{-3}}} \right)    \left( \frac{v}{1000 {\rm km/s}} \right)   \left( \frac{R}{500{\rm kpc}} \right)     \left( \frac{\Delta t}{10^{9} {\rm yr}} \right)   \left(\frac{m}{10^{11} M_{\rm \odot}}\right)^{-1}  {\rm kpc},
\end{equation}
where $n_{\rm ICM}$ is the ICM number density. 
Contrary to the dynamical friction (Eq.6) which is more effective on more massive galaxies, the ram pressure
can thus drag less massive ones more effectively. 
In addition, the ICM can also affect the galaxy motion via viscosity. As shown in, e.g., Nulsen (1982), the viscous force 
by the laminar ICM flow through a galaxy can be written as
\begin{equation}
F_{\rm VIS}(R) \approx \pi R_{\rm int}^2 \rho_{\rm ICM}(R) v^2 12/ R_{\rm e}, 
\end{equation} 
where $R_{\rm e}$ is the Reynolds number of the ICM. This has the same form as Eq.7, and its effect would be comparable to the ram pressure 
when the Reynolds number of ICM is low.

The influence of the ICM ram pressure on the stellar component of galaxies has been extensively investigated
in the last decade (e.g., Yoshida et al. 2004, Kenney et al. 2004). By analyzing the combined UV to radio data
of the Virgo galaxies, Boselli et al. (2009) discovered that the ram pressure would be responsible for morphological 
disturbances of both the gas and the young stellar population in the disk. The underlying physics may be revealed in
the simulation work of Vollmer (2003) and Steinhauser et al. (2012), that the stars formed in the ram-pressure tails can interact
gravitationally with the parent galaxy and eventually make its disk thicker. They also discovered that, when exposed to a medium-level ram 
pressure, the interstellar medium would not be immediately stripped, but remains displaced to the downstream direction (i.e., backward) on a timescale of 
several hundred Myr, to gravitationally pull the entire galaxy backwards. In such a way, the ICM ram pressure can 
induce drag on not only the gaseous component, but also
on the other mass components (stars and dark matter, in particular) of a galaxy.

As shown in Eq.7, the strength of galaxy-ICM interaction is proportional to the galaxy velocity squared and the ambient 
gas density. Since the galaxy velocity measurements are less complete with the current data (\S3.1.3), it is natural to examine whether 
the GNIMR evolution depends on the ICM density. Here we calculated the ICM density at $r=0.5 R_{500}$
for each cluster using the X-ray spectroscopy results (\S3.2), and divided each subsample into two parts by a dividing value 
$\rho^{\ast}_{\rm ICM} = 10^{-3} {\rm cm^{-3}}$. As shown in Figure 22a and 22b, in a high-$\rho_{\rm ICM}$ environment, galaxies have 
in fact been concentrated to within $\leq 0.1$ $R_{500}$, whereas such an evolution in galaxy distribution is weaker in low-$\rho_{\rm ICM}$
clusters. As a result, the GNIMR profiles of the high-$\rho_{\rm ICM}$ objects (Fig. 22d) exhibit nearly the same pattern of evolution as the sample-average 
one shown in Figure 14b, while the low-$\rho_{\rm ICM}$ profiles (Fig. 22c) suggest a less significant evolution. 
Therefore, in $z < 0.5$, the clusters
with relatively lower ICM densities have a slightly smaller number of galaxies infalling towards the center than those 
with higher ICM densities. These results support the prediction in Paper I, that the galaxy-ICM 
interaction contributes significantly to the observed GNIMR evolution.
On the other hand, for the subsample H, the average 
GNIMR profile of the relatively low-$\rho_{\rm ICM}$ clusters appears to be 
as steep as the one of high-$\rho_{\rm ICM}$ objects, indicating that the galaxy-to-ICM concentration was already
in place at $z \sim 0.5$ for the entire $\rho_{\rm ICM}$ range considered in this work.

The above scenario describes the interaction between individual galaxies and its environment. In reality, galaxies are not always travelling
alone; a fraction of infalling galaxies are bound to subcluster-scale groups before fully merged with the cluster. 
In such a group-in-cluster configuration, both the galaxy-cluster and group-cluster interactions are naturally expected. Based
on a weak lensing study in the central 1.6 Mpc region of the Coma cluster, Okabe et al. (2014) discovered 32 subcluster-scale halos,
 each with a mass of several times $10^{12}$  
$M_{\odot}$ to $10^{13}$ $M_{\odot}$. According to Eq.6, the dynamical
friction can affect the clustocentric distances of these halos by $50-150$ kpc per Gyr. Meanwhile, Sanders et al.
(2013) observed the same cluster in X-rays, and discovered two coherent linear ICM structures up to 650 kpc. These
structures provide smoking-gun evidence for strong direct interactions, e.g., ram pressure, between infalling subclusters and cluster ICM 
on several Myr. The interaction between the ICM and individual galaxies in the group may also be enhanced due to a large relative
velocity $v$. As reported in Yagi et al. (2015), most of the galaxies showing features of strong interaction, e.g., extended H$\alpha$ emission, in two $z=0.4$
clusters are found to belong to infalling subclusters. In addition, in such subclusters, pairs of galaxies orbiting each other will
lose their angular momentum, via interaction with the cluster environment, and will merge together on the infalling timescales. Thus,
the interaction is considered to enhance morphological evolution of member galaxies.

\subsection{An Energy Flow on Cosmological Timescales}

The observed persistent and large-scale matter infall, first predicted by M01 and confirmed through Paper I and the present study, is expected to generate a large energy flow from galaxies
to the ICM. The energy loss rate of a single galaxy in the interactions can be represented by $L = (F_{\rm DF} + F_{\rm RP}) \times v$.
Employing typical values, the energy flow per galaxy can be written as 
\begin{equation}
\begin{array}{rcl}
L \approx 1 \times 10^{41}   \left( \frac{\rho}{10^{-4} M_{\rm \odot}{\rm pc}^{-3}} \right)   \left( \frac{m}{10^{11} M_{\rm \odot}} \right)^{2}  \left( \frac{v}{1000 {\rm km/s}} \right)^{-2}  \\
+ 2 \times 10^{41} \left( \frac{R_{\rm int}}{5{\rm kpc}} \right)^{2}  \left( \frac{n_{\rm ICM}}{10^{-3} {\rm cm^{-3}}} \right)   \left( \frac{v}{1000 {\rm km/s}} \right)^{3} {\rm erg s^{-1}}.
\end{array}
\end{equation}
The energy thus transfers silently from thousands of galaxies to the environment, creating a flow of $10^{44-45}$ erg s$^{-1}$ per
cluster on several Gyr. This makes it one of the largest energy events in the Universe, comparable to outbursts of most luminous AGNs, and to the X-ray luminosity 
from each cluster. Furthermore, the spatial distribution of 
$L$ is highly concentrated in cluster center, where the galaxy-ICM term $F_{\rm RP}$ has a larger contribution than gravitational term $F_{\rm DF}$
due to larger galaxy velocities therein. As a result, more than half of the energy is expected to transfer directly to the ICM component,
which significantly contributes to the ICM heating to suppress cooling flows (M01). Furthermore, this can explain how the ICM component has
been evolving to achieve larger angular extents than the galaxy and dark matter components in nearby clusters (Fig. 14f). Such a gas heating by galaxy infall has been successfully 
reproduced with recent numerical simulations (e.g., Asai et al. 2007, Ruszkowski \& Oh 2011, Parrish et al. 2012). 

Besides the possible ICM heating, the current scenario has several other important implications. First, it immediately explains how the centrally
concentrated galaxy distributions in nearby clusters were formed. Second, it provides an important clue to the origin of the environmental effects
and evolution seen in member galaxies (Butcher \& Oemler 1984; Wen \& Han 2015a). Third, it may potentially explain the observed central drop in IMLR, and the large-scale ($\geq R_{\rm vir}$) abundance uniformity 
in the ICM (Werner et al. 2013).

\section{CONCLUSION}

In Paper I, we measured radial profiles of stellar light and ICM mass for 34 galaxy clusters with redshifts
of $0.1-0.9$, and detected, for the first time, a significant evolution that the member galaxies get more 
centrally-concentrated with time relative to the ICM and dark matter. By using the SDSS photometric and 
{\it Chandra}/{\it XMM-Newton} data, now we have greatly enhanced
our study by constructing an unprecedented cluster catalog with 340 X-ray bright clusters, although the redshift range
($z < 0.5$) became somewhat narrower. Using this new sample,
we have quantitatively confirmed and reinforced the results reported in Paper I. While the member
galaxies are continuously falling to the center relative to the ICM and dark matter, the ICM has been
slightly expanded relative to the dark matter even though it keeps radiating. The observed evolution cannot
be explained by various systematic errors or z-dependent selection biases. The galaxy infall is seen both in
more massive and smaller member galaxies, and is enhanced in clusters with higher ICM densities. Therefore, the
observed effects cannot be explained by dynamical friction alone, but require more direct galaxy-ICM interaction. 
These interactions are considered to create a large energy flow of $10^{44-45}$ erg $\rm s^{-1}$ per cluster from
the member galaxies to their environment, which is estimated to continue over cosmological timescales. The present results
have several important implications for the evolution of clusters and galaxies within them.

\section*{Acknowledgments}
We thank Jinlin Han, Masamune Oguri, Kazuhiro Shimasaku, Yuanyuan Su for helpful suggestions and comments. This work was supported by the
Ministry of Science and Technology of China (grant No. 2013CB837900), the National
Science Foundation of China (grant Nos. 11125313, and 11433002).

\clearpage

\begin{deluxetable}{lccccccccc}
\centering
\tablecaption{Sample Statistics
\label{tbl:Genlog} }\tablewidth{0pt} \tablecolumns{6} 
\tablehead{
 & \colhead{raw} & \colhead{selected} &
\colhead{SDSS} & \colhead{X-ray quality}\tablenotemark{a} & \colhead{$M_{500}$ filter} }

\startdata

{\it Chandra} & 509 & 332 & 223 & 161 & 141 \\ 
{\it XMM-Newton} & 442 & 357 & 245 & 220 & 199 \\ 
Total & 951 & 689 & 468 & 381 & 340 \\

\enddata
\scriptsize
\tablenotetext{a}{The neighboring Virgo, Perseus, and Coma clusters are also removed.}

\end{deluxetable}

\clearpage

\begin{figure}
\begin{center}
\includegraphics[angle=-0,scale=.5]{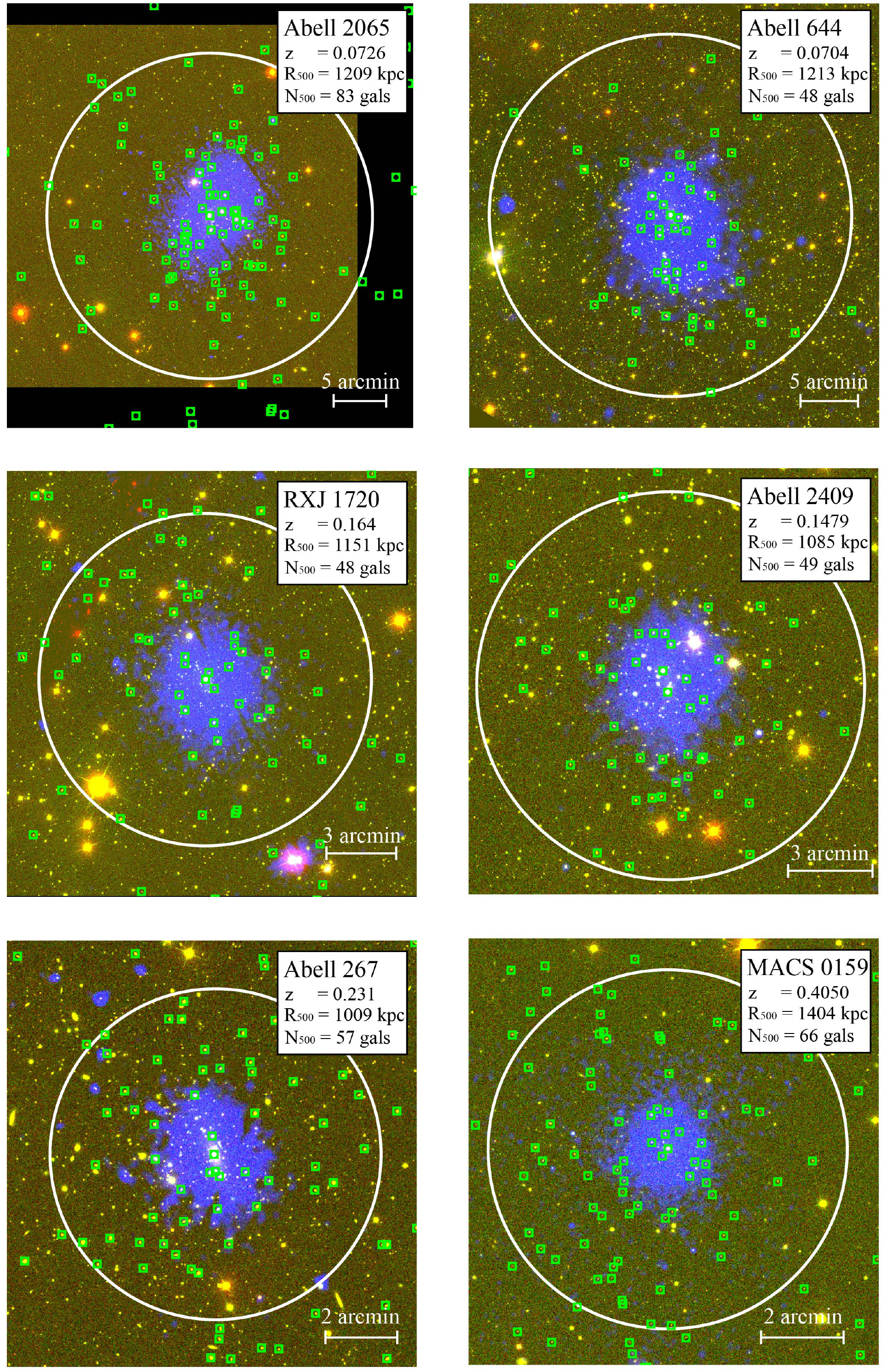}
\caption{RGB-colored composite images (red for SDSS-i, green for SDSS-g, and blue for {\it Chandra}/{\it XMM-Newton} $0.3-0.8$ keV) of
six example clusters in our sample. Detected member galaxies (\S2.4) are marked by small green boxes. X-ray determined $R_{500}$ of each cluster is shown with
a white circle. }
\end{center}
\end{figure}
\clearpage

\begin{figure}
\begin{center}
\includegraphics[angle=-0,scale=.4]{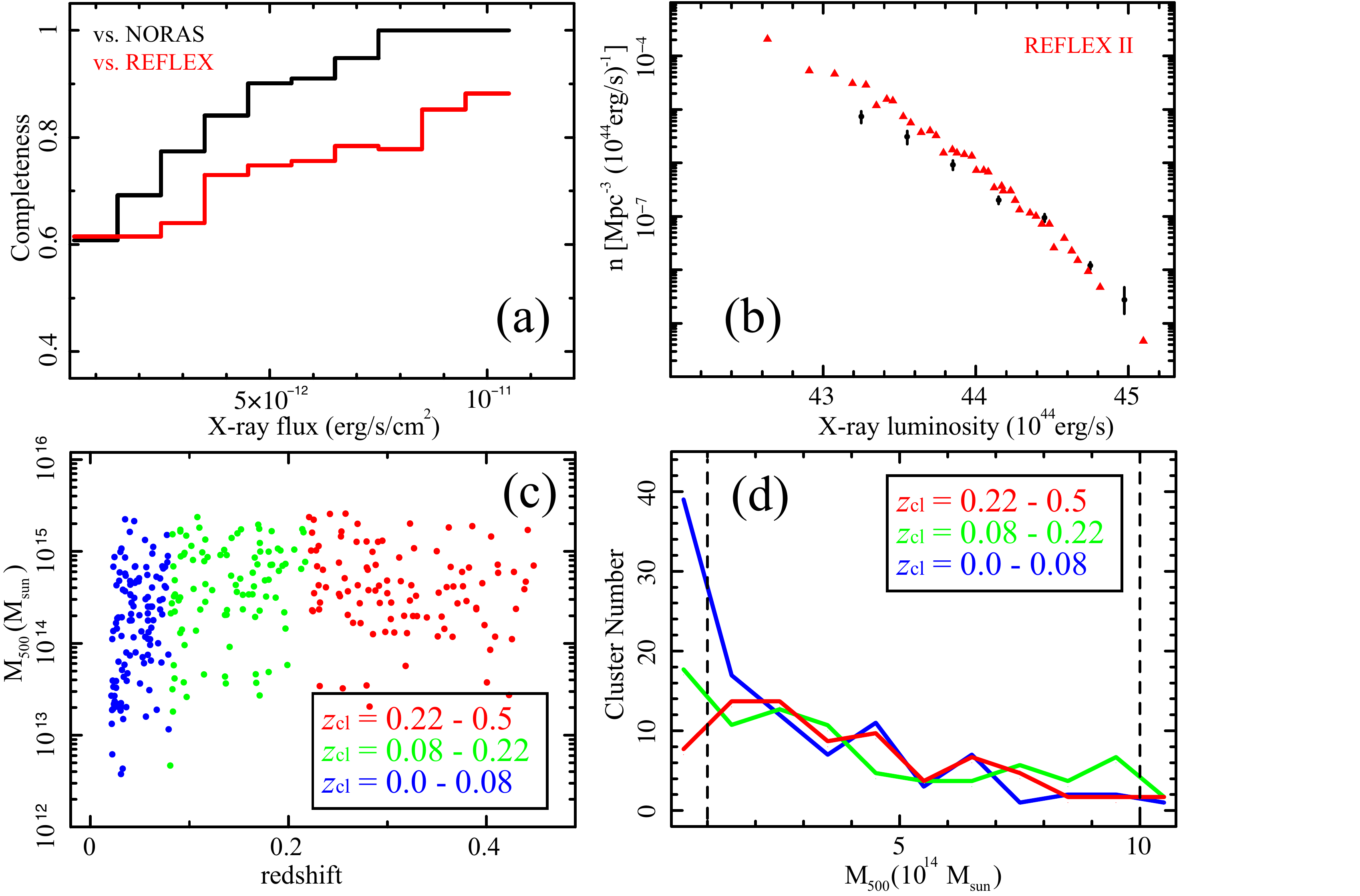}
\caption{(a) Completeness of the Intermediate Sample compared to the existing X-ray-selected cluster
catalogs (NORAS and REFLEX in black and red, respectively). (b) Comparison between the 
X-ray luminosity functions of the sample (black) and that of the REFLEX II sample (red; B$\rm \ddot{o}$hringer 
et al. 2014). (c) A plot of $M_{500}$ vs. redshift of the sample. Subsamples L, M, and H are 
shown with blue, green and red points, respectively. (d) Cluster mass functions of the 
three subsamples. Two vertical lines are the limiting values used in this work.   }
\end{center}
\end{figure}
\clearpage

\begin{figure}
\begin{center}
\includegraphics[angle=-0,scale=.33]{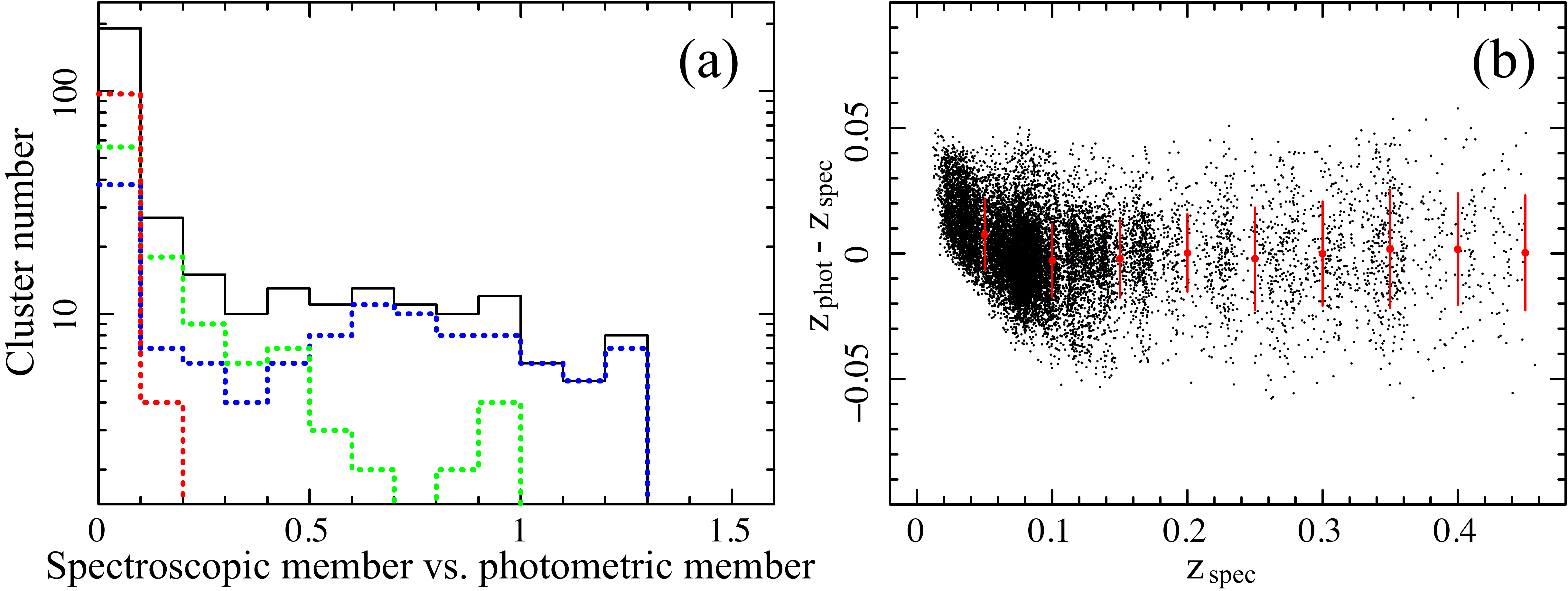}
\caption{(a) Histograms of the sample clusters, plotted as a function of the ratio between the members selected 
by $z_{\rm spec}$ and those selected by $z_{\rm phot}$. Subsamples L, M, and H are plotted with blue, green
and red dashed lines, respectively, and the total is shown in black solid line. (b) Comparison between
$z_{\rm spec}$ and $z_{\rm phot}$ for all the member galaxies that have both spectroscopic and photometric measurements. 
The average value and scatter in each redshift bin is shown with red error bars.}
\end{center}
\end{figure}
\clearpage

\begin{figure}
\begin{center}
\includegraphics[angle=-0,scale=.4]{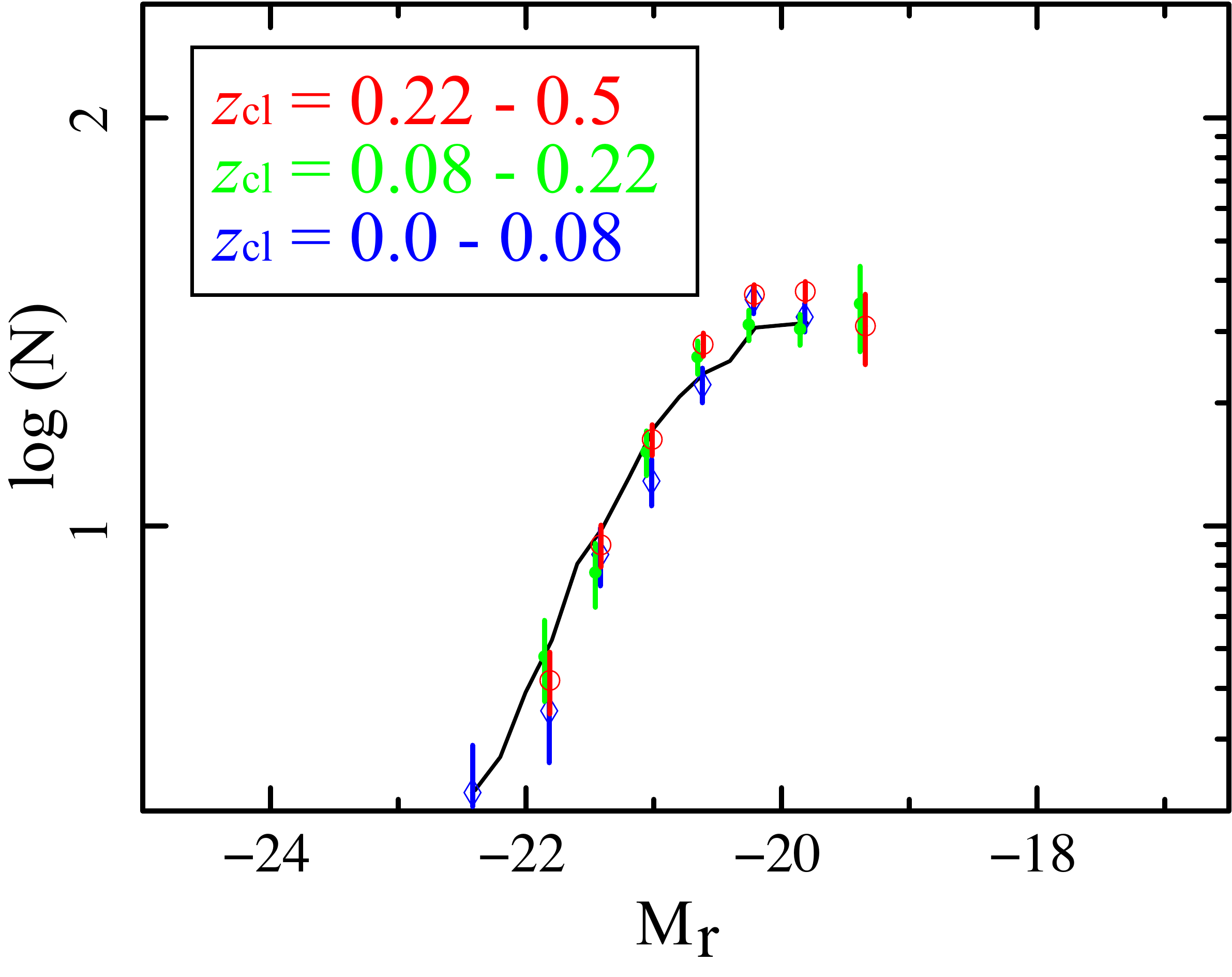}
\caption{$r$-band luminosity functions of the identified member galaxies of the Final Sample. The subsamples L, M, and H are
  plotted in blue, green and red, respectively, shown after several corrections described in the text. The solid line is a reference luminosity function
reported in Wen \& Han (2015b) for a more extensive SDSS sample at $z = 0.05 - 0.15$.}
\end{center}
\end{figure}
\clearpage

\begin{figure}
\begin{center}
\includegraphics[angle=-0,scale=.4]{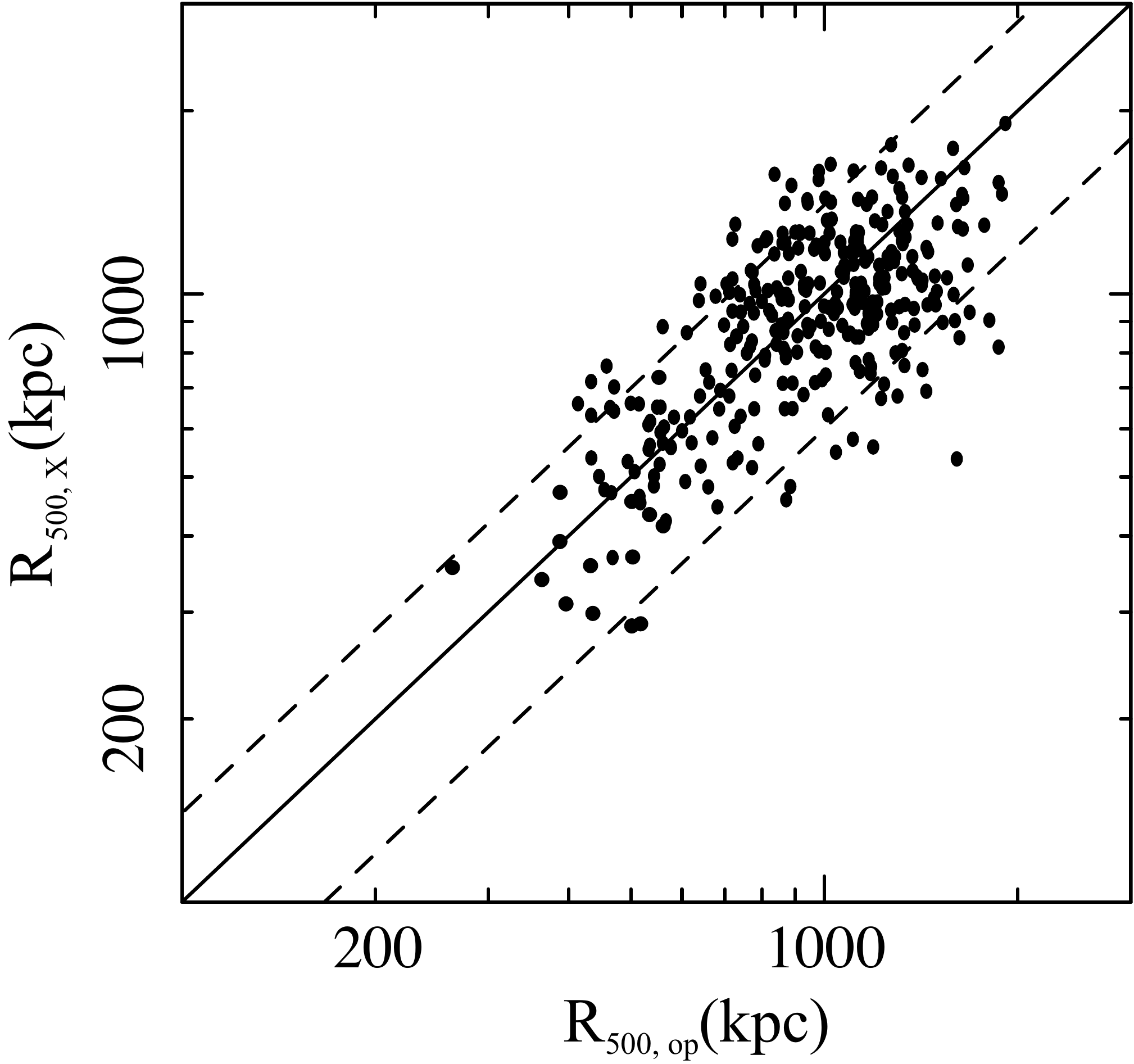}
\caption{Comparison between the values of $R_{500}$ of the Final Sample clusters derived with the X-ray hydrostatic estimate (ordinate) and those with
a scaling of total optical luminosity via Eq.1 (abscissa). Solid line means an exact agreement, while the two dash lines 
indicate 30\% discrepancies.}
\end{center}
\end{figure}
\clearpage

\begin{figure}
\begin{center}
\includegraphics[angle=-0,scale=.4]{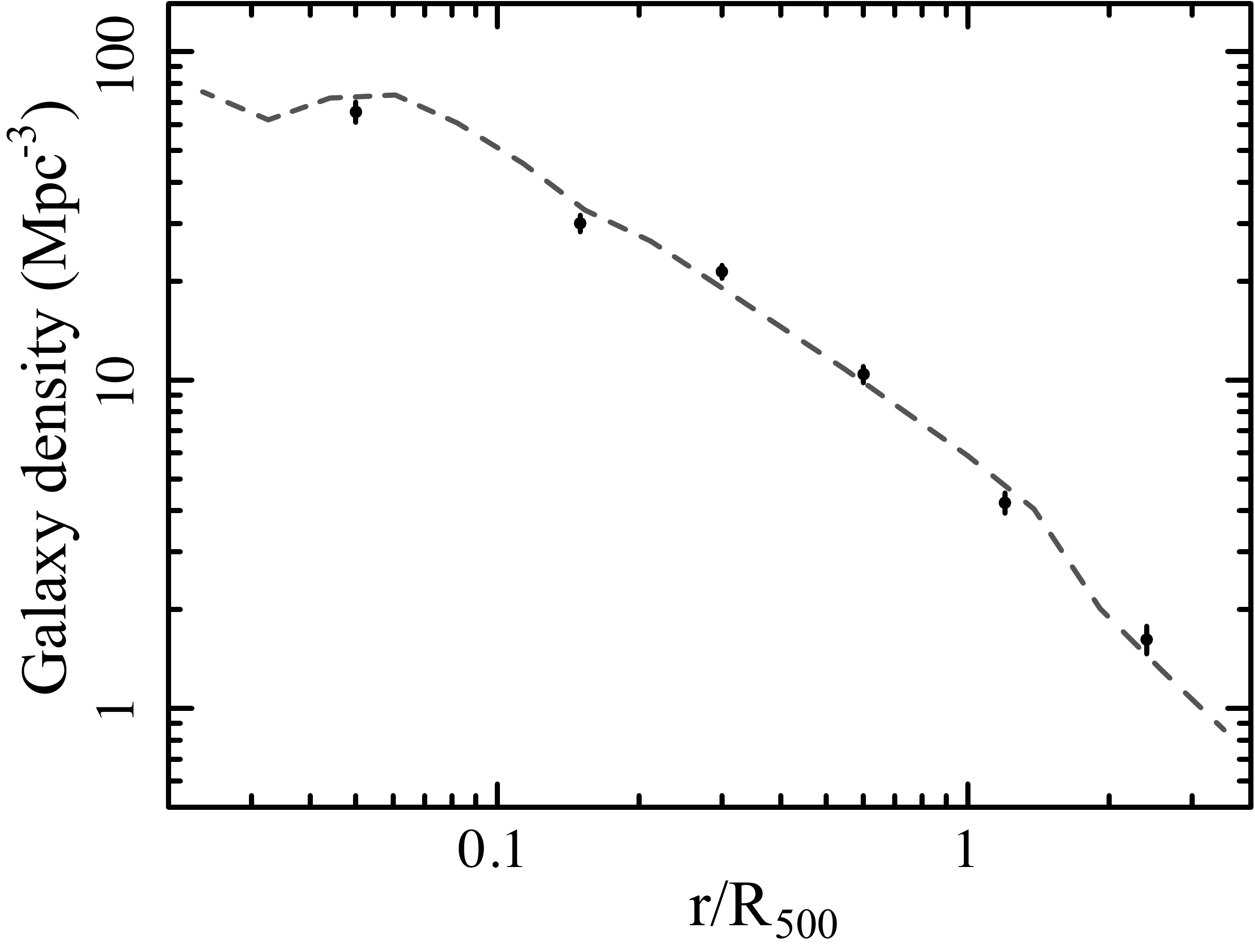}
\caption{The galaxy number density profile averaged over the entire Final Sample, with 68\% error bar. The dashed line is the mean
galaxy density profile from Budzynski et al. (2012). Both profiles are scaled to $R_{500}$.}
\end{center}
\end{figure}
\clearpage

\begin{figure}
\begin{center}
\includegraphics[angle=-0,scale=.25]{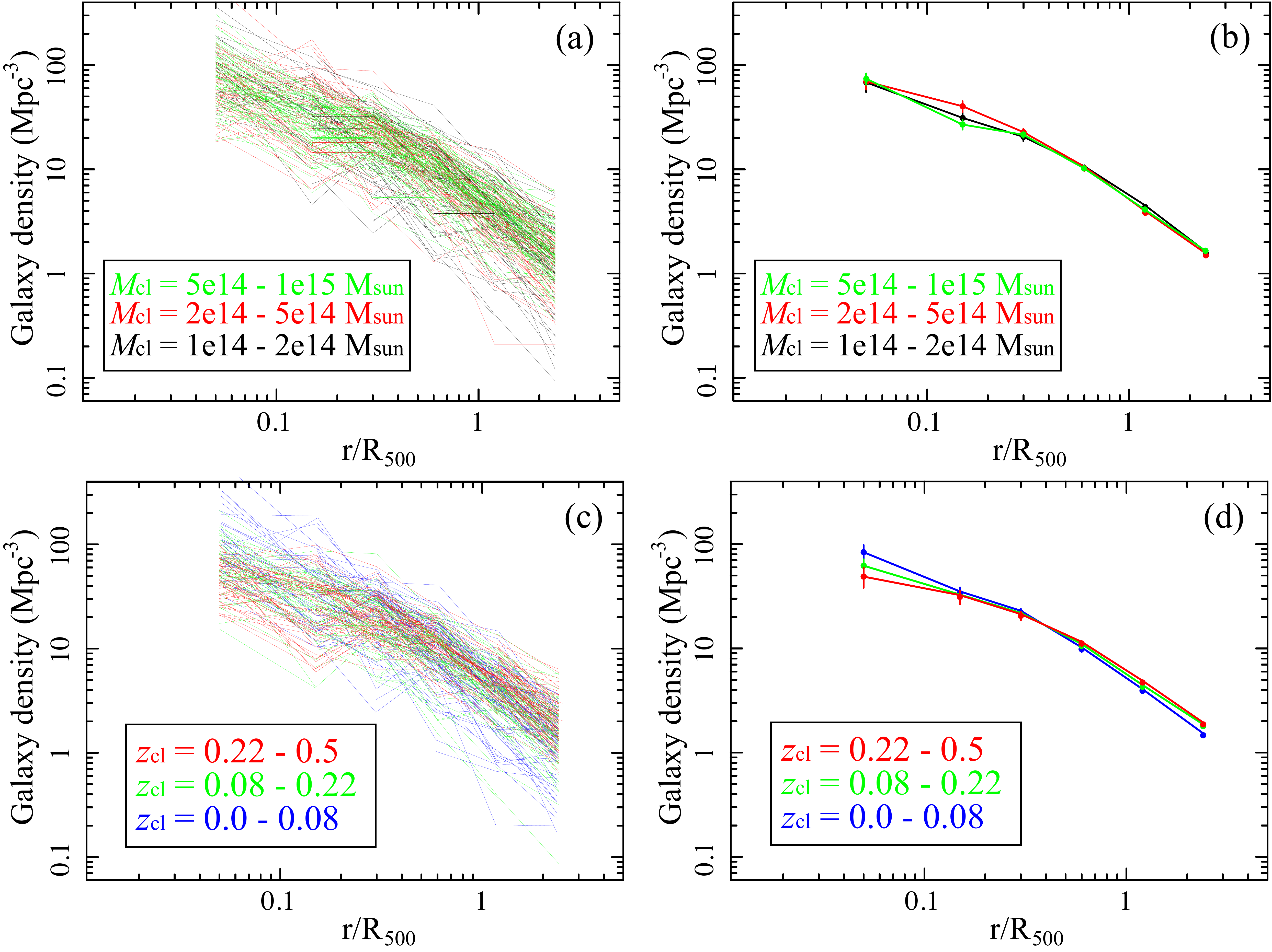}
\caption{(a) Galaxy density profiles of all sample clusters, divided into three groups by
the cluster mass. The low mass, intermediate mass, and high mass systems are presented in black, red,
and green. (b) The mean profiles of the three mass-bins. (c) The same as panel (a), but the color specifies
the redshift subsamples. The subsample L, M, and H are shown with blue, green, and red, respectively. (d) The mean profile of
the three subsamples. }
\end{center}
\end{figure}
\clearpage

\begin{figure}
\begin{center}
\includegraphics[angle=-0,scale=.4]{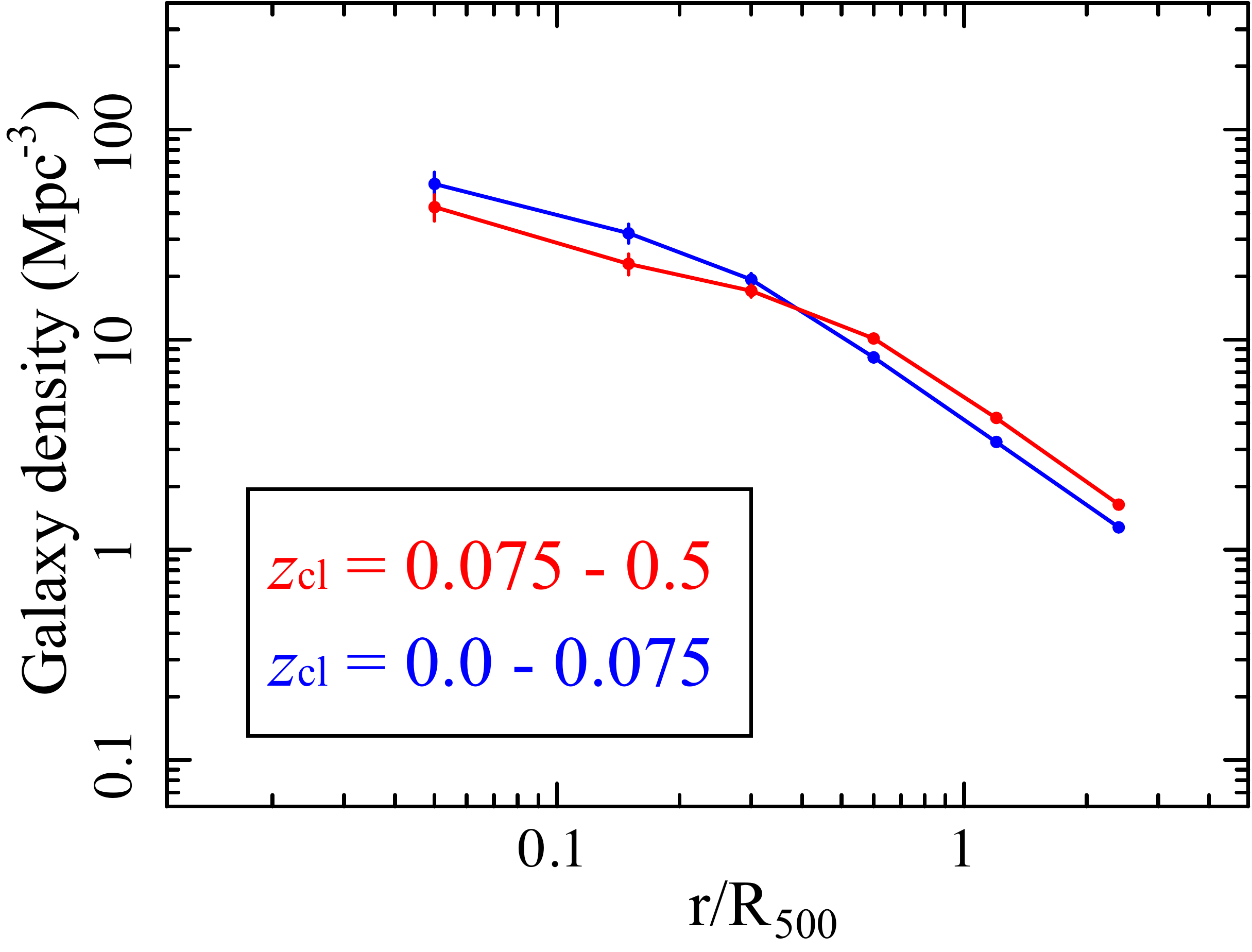}
\caption{Same as Figure 7d, but using only the spectroscopically measured galaxies, and dividing the
clusters into two redshift-dependent subsamples instead of the three. }
\end{center}
\end{figure}
\clearpage

\begin{figure}
\begin{center}
\includegraphics[angle=-0,scale=.4]{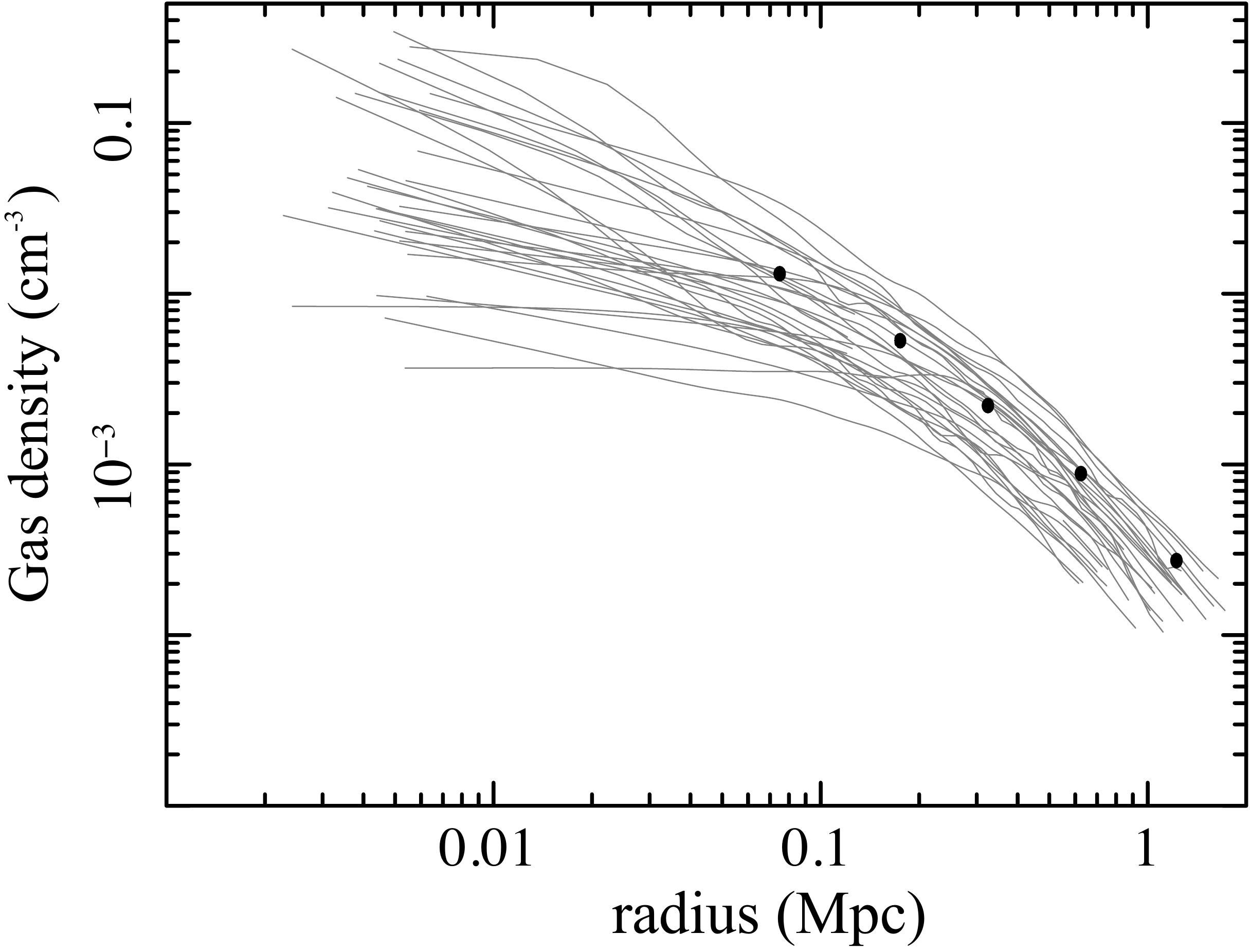}
\caption{The mean ICM density profile of the our sample clusters (black data points). The 68\% error is also plotted,
although it is too small to be visually apparent. The reference ICM density profiles
obtained by Croston et al. (2008) are shown in grey thin lines. }
\end{center}
\end{figure}
\clearpage

\begin{figure}
\begin{center}
\includegraphics[angle=-0,scale=.33]{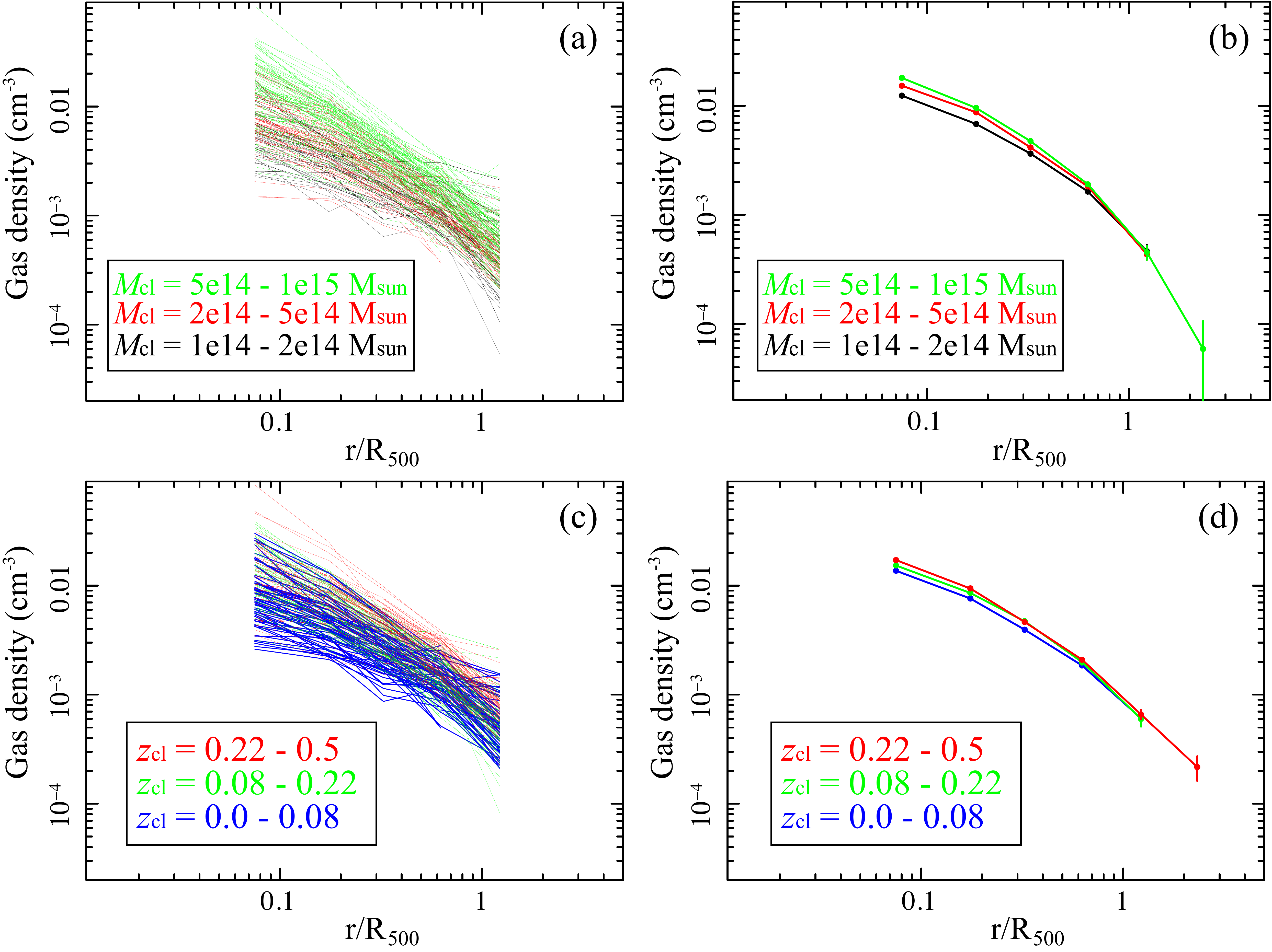}
\caption{(a) ICM density profiles of all clusters in the sample. The black, red, and green curves
are low-mass, intermediate-mass, and high-mass systems, respectively. All profiles are scaled to 
the characteristic radius $R_{500}$. (b) Mean gas density of the 
three mass groups. (c) The same as panel (a), but the color specifies the redshift subsamples. The subsample L, M, and H are shown in 
blue, green, and red, respectively. (d) Subsample-averaged ICM density profiles. }
\end{center}
\end{figure}
\clearpage

\begin{figure}
\begin{center}
\includegraphics[angle=-0,scale=.4]{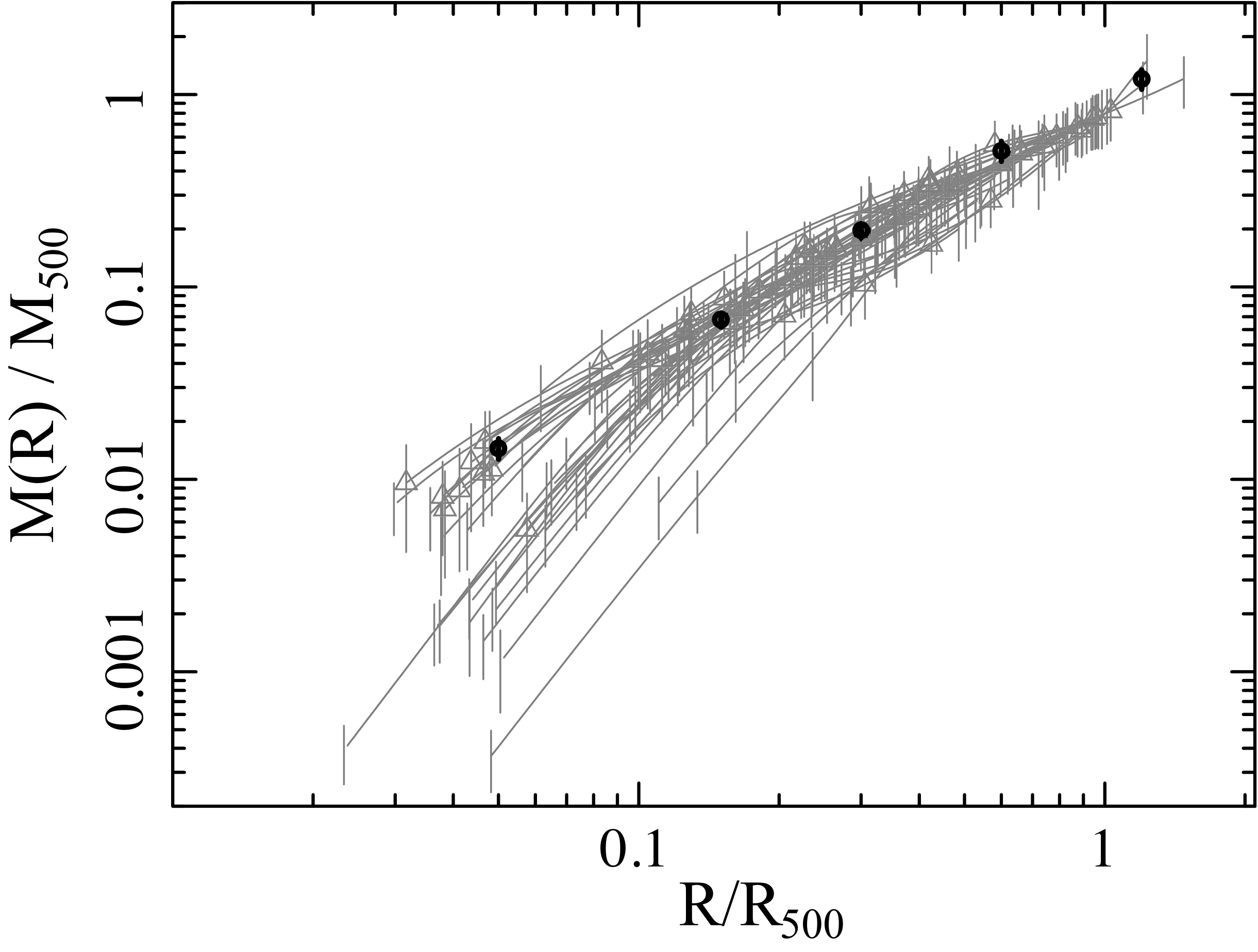}
\caption{Mean radially integrated profile of the gravitating mass distributions of the sample (black data points). The reference
mass profiles presented in Zhang et al. (2008) are plotted in grey error bars and thin lines. }
\end{center}
\end{figure}
\clearpage


\begin{figure}
\begin{center}
\includegraphics[angle=-0,scale=.33]{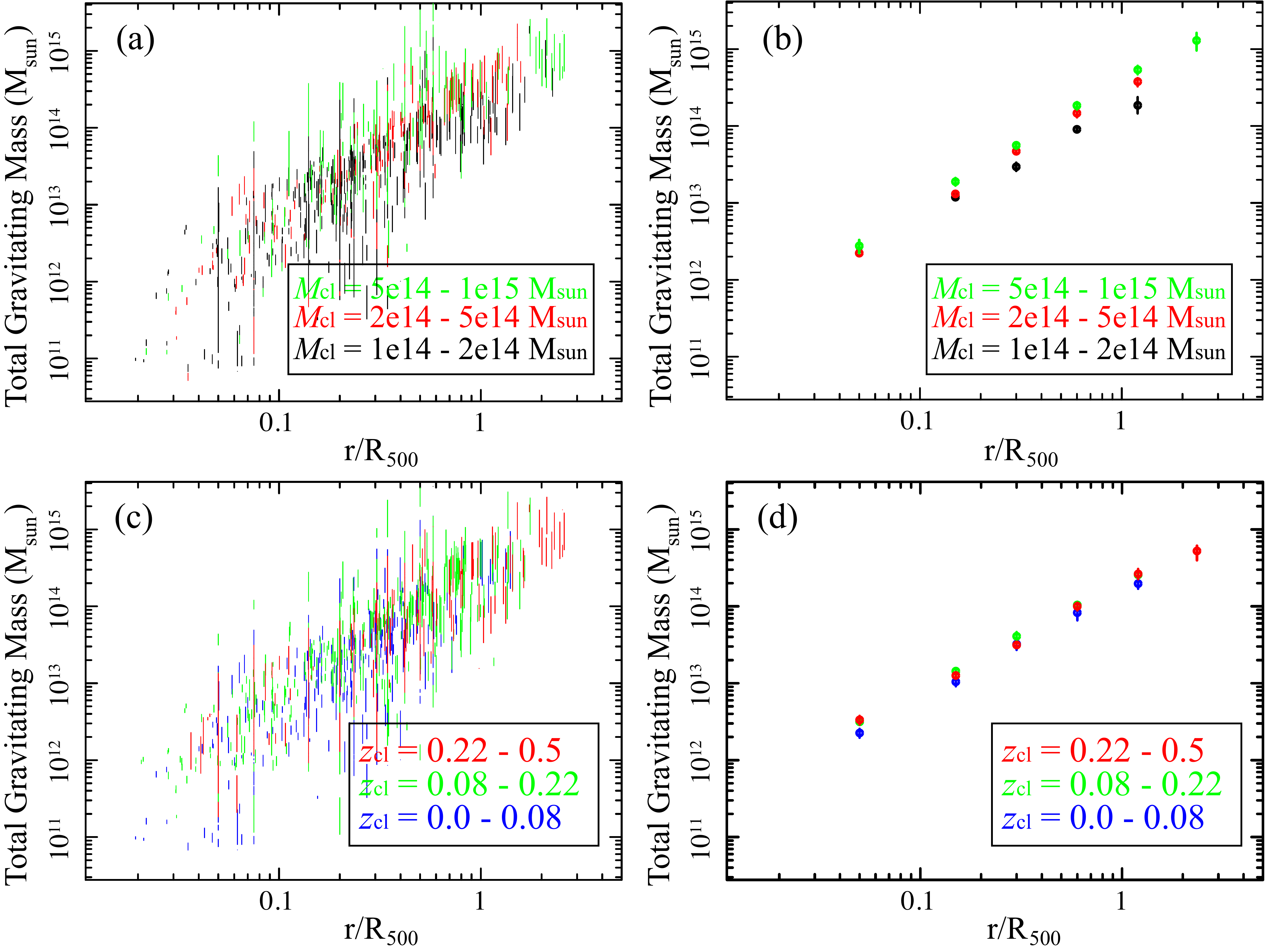}
\caption{(a) Gravitating mass profiles of the Final Sample clusters. The black, red, and green 
points specify the low-mass, intermediate-mass, and high-mass ranges, respectively. The radii are scaled
to the characteristic value $R_{500}$. (b) Mean mass profiles for the three mass ranges. 
(c) The same as (a), but the subgrouping is based on the redshift; subsample L, M, and H are represented
in blue, green, and red, respectively. (d) Mean mass profiles for the three subsamples. }
\end{center}
\end{figure}
\clearpage

\begin{figure}
\begin{center}
\includegraphics[angle=-0,scale=.4]{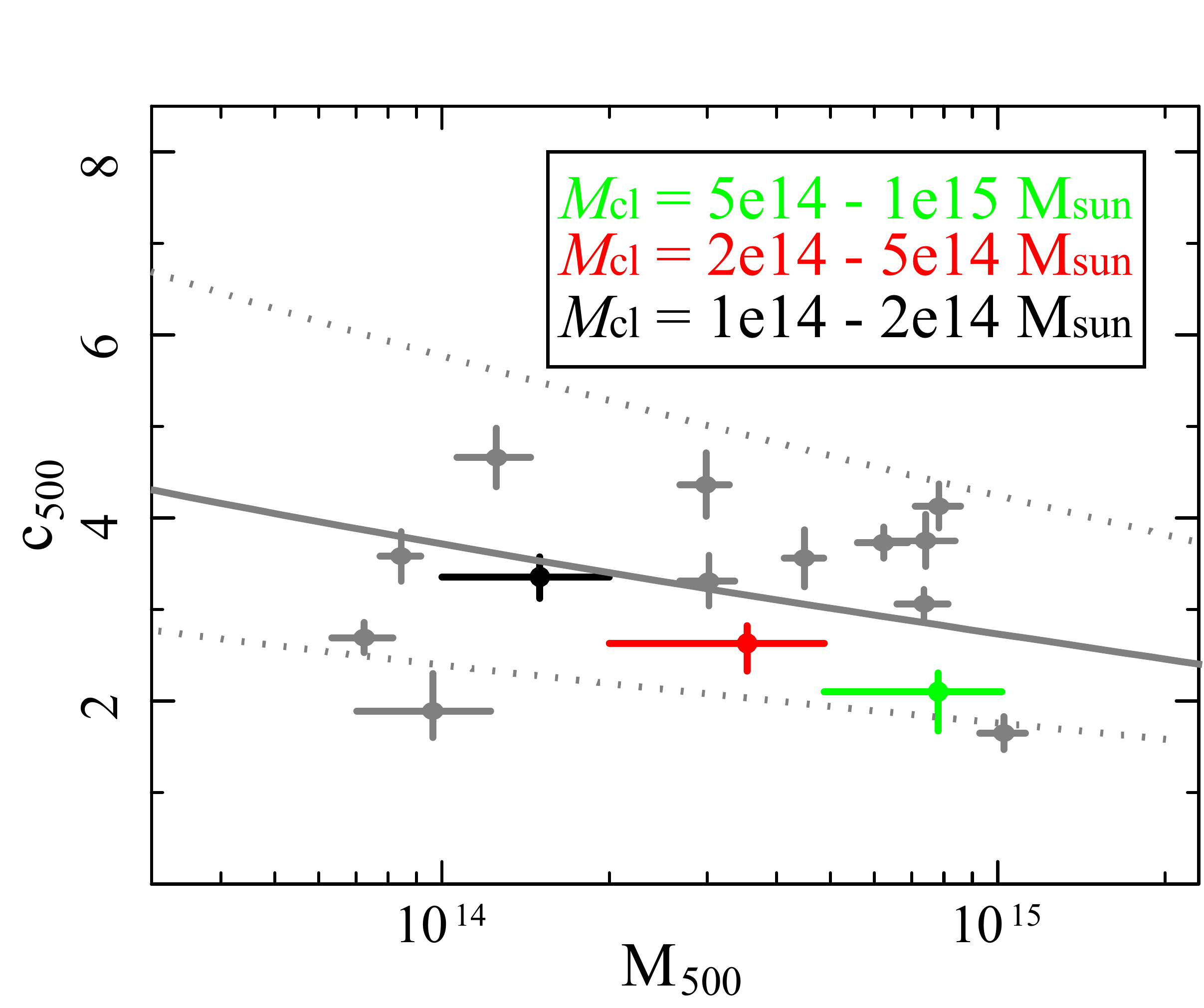}
\caption{The concentration parameters of the NFW model, $c_{500} = R_{500}/ R_{\rm s}$, as a function
of the cluster mass $M_{500}$. The black, red, and green points are the mean values for the low-mass,
intermediate-mass, and high-mass clusters, respectively. The grey points are those measured in X-rays by
Vikhlinin et al. (2006). Solid line is the $c_{500}-M_{500}$ relation from a numerical simulation
by Dolag et al. (2004), and the dotted lines specify the 2-$\sigma$ scatter around the simulated relation. }
\end{center}
\end{figure}
\clearpage

\begin{figure}
\begin{center}
\includegraphics[angle=-0,scale=.25]{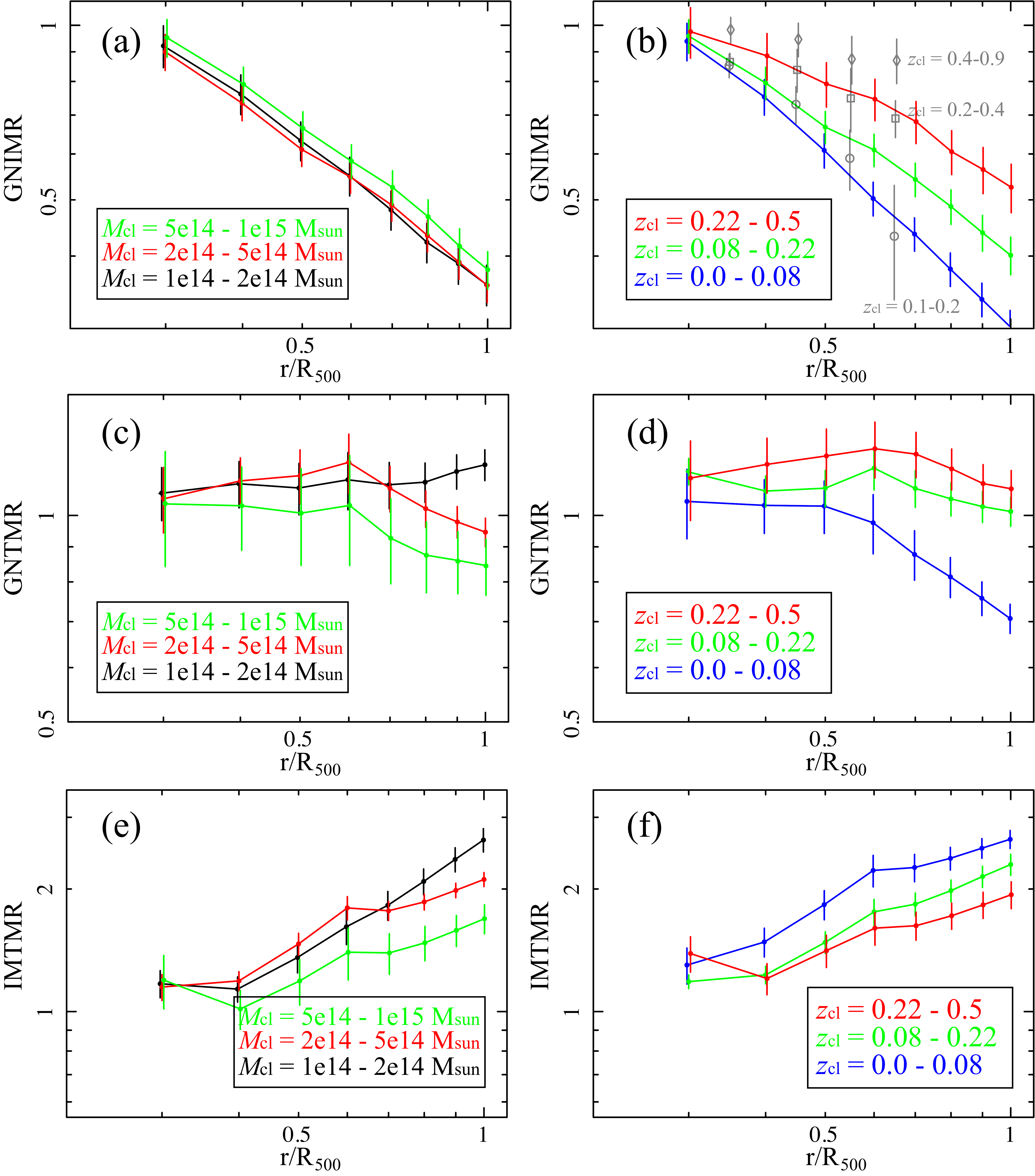}
\caption{(a) Mean galaxy number vs. ICM mass ratio (GNIMR) profiles of the low-mass clusters (black),
intermediate-mass clusters (red), and high-mass clusters (green). (b) Mean GNIMR profiles of clusters
in the subsample L (blue), M (green), and H (red). Grey points and error bars show the GNIMR profiles
measured in Paper I. (c) Mean galaxy number vs. total mass ratio (GNTMR) profiles of the three mass groups. 
(d) Mean GNTMR profiles of the three redshift subsamples. (e) Mean ICM mass vs. total mass ratio (IMTMR) profiles
of the three mass bins. (f) Mean IMTMR profiles of the three subsamples.}
\end{center}
\end{figure}

\clearpage

\begin{figure}
\begin{center}
\includegraphics[angle=-0,scale=.4]{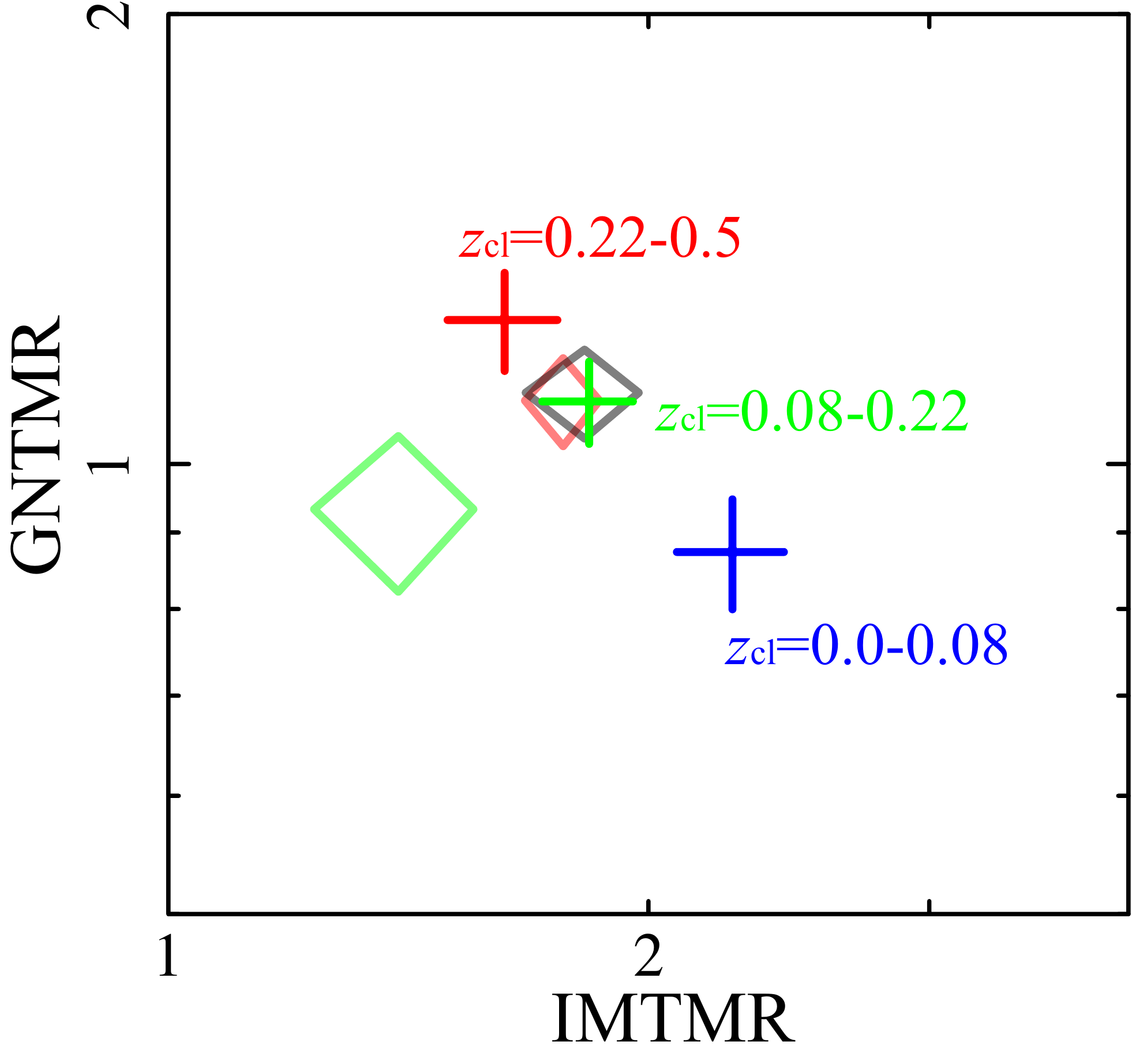}
\caption{Characterization of the clusters on the GNTMR vs. IMTMR plane. The data points with cross
are the mean values of the three redshift-sorted subsamples (subsample L, M, and H in blue, green, and red, 
respectively). The mean values of the three mass-sorted subsamples are shown in diamonds, and the 
black, red, and green stand for low-mass, intermediate-mass, and high-mass bins, respectively. }
\end{center}
\end{figure}

\clearpage

\begin{figure}
\begin{center}
\includegraphics[angle=-0,scale=.4]{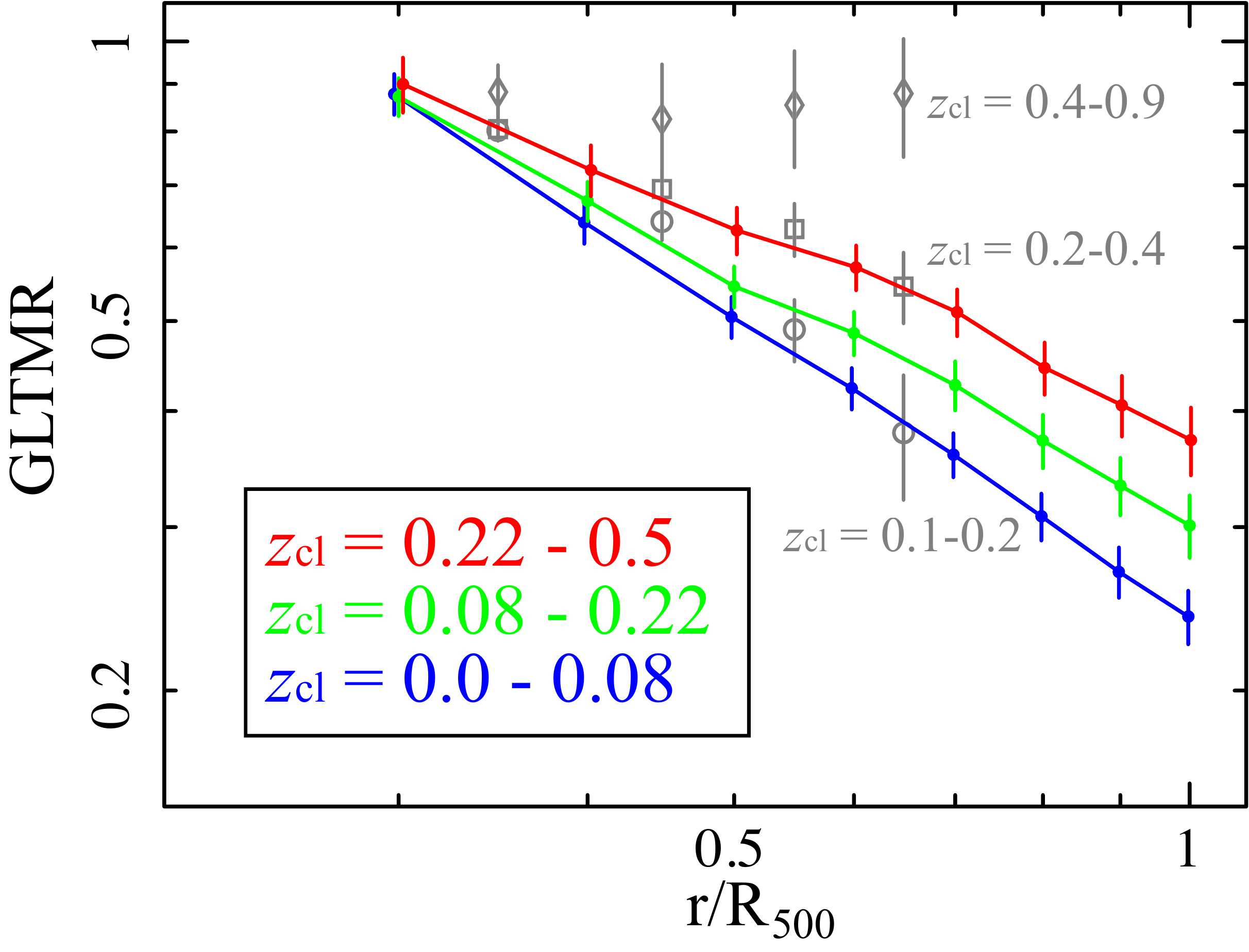}
\caption{Galaxy light vs. ICM mass ratio (GLIMR) profiles averaged over the subsample L, M, and H, 
  shown in blue,
  green, and red colors, respectively. As a reference, the GLIMR results from 
Paper I are given in open grey symbols. }
\end{center}
\end{figure}

\clearpage

\begin{figure}
\begin{center}
\includegraphics[angle=-0,scale=.22]{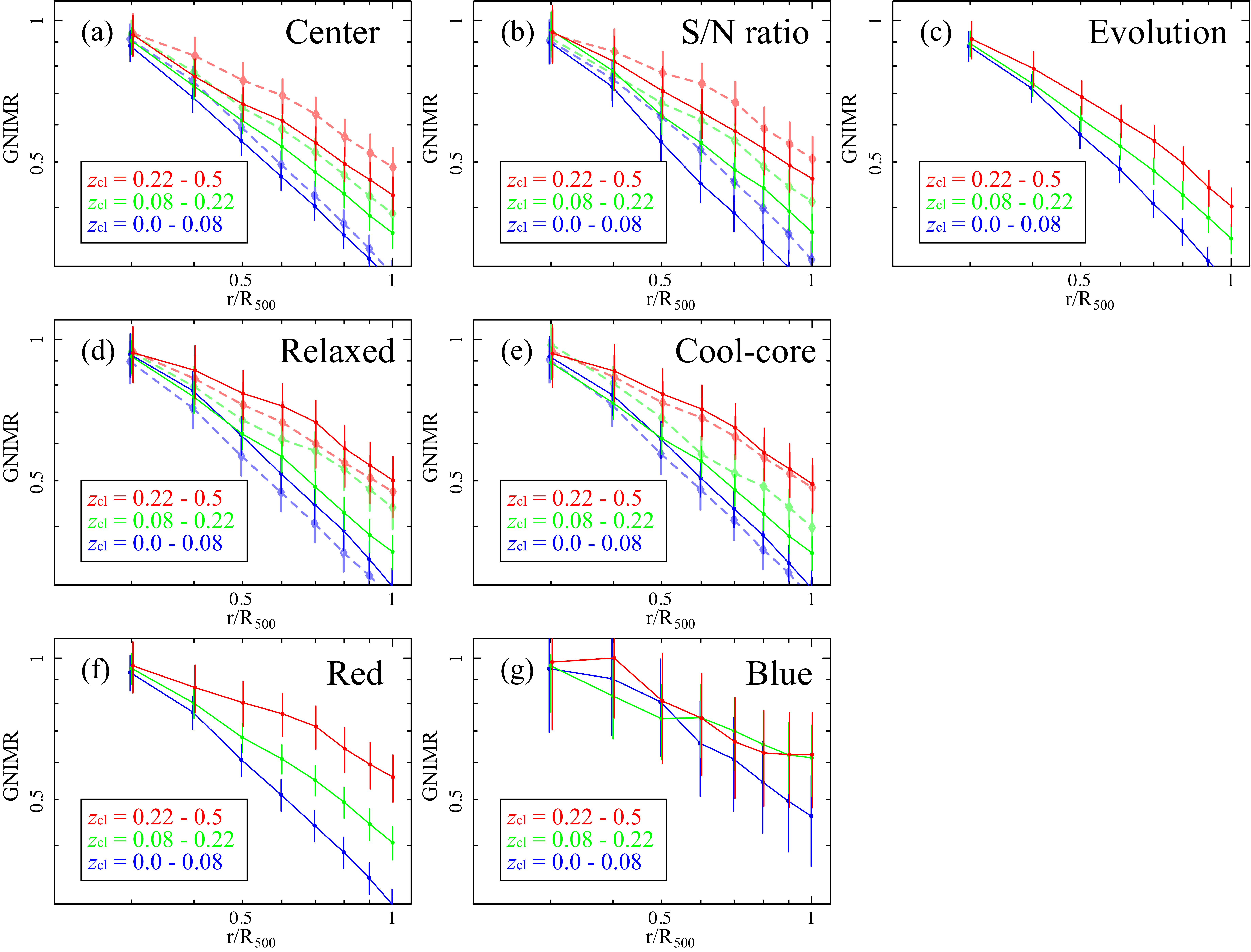}
\caption{Effects of various biases on the subsample-averaged GNIMR profiles. 
The subsample L, M, and H are plotted in blue, green, and red, respectively. 
(a) Subsample-averaged GNIMR profiles calculated with optically-defined 
center (solid lines) and those with X-ray-defined center (dashed fainter lines). The
difference of the centers applies only to the galaxy profiles, while the ICM profiles 
are the same. (b) Those of clusters
with high (solid lines) and low (dashed lines) X-ray signal-to-background ratio. 
(c) Those calculated based on the predicted evolutionary scale $R_{500}^{z=0}$. 
(d) Comparison of the 
clusters with relaxed optical morphology ($\Gamma > 0$;
solid lines) and those with unrelaxed morphology ($\Gamma < 0$; dashed fainter lines). (e) Comparison between 
the clusters with a likely cool-core (solid lines) and those without a cool-core (dashed fainter lines). 
(f) Results using only selected red galaxies. (g) Those obtained with 
only blue galaxies. }
\end{center}
\end{figure}

\clearpage

\begin{figure}
\begin{center}
\includegraphics[angle=-0,scale=.4]{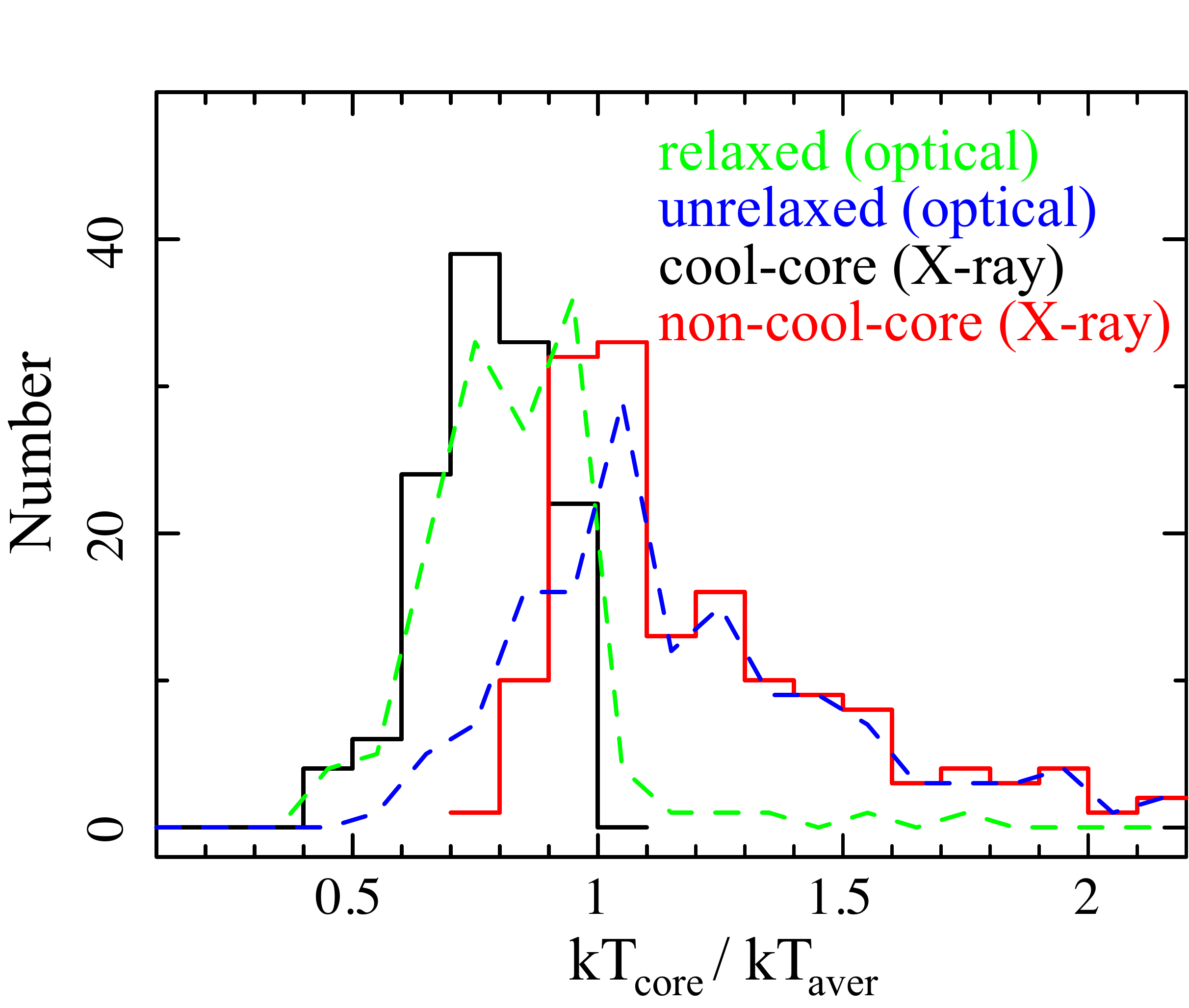}
\caption{Histograms of the morphologically relaxed clusters (green) and unrelaxed ones (blue),
as a function of the ratio between the core ICM temperature and the cluster average temperature. The black
and red histograms show the cool-core and non-cool-core systems, respectively. }
\end{center}
\end{figure}

\clearpage

\begin{figure}
\begin{center}
\includegraphics[angle=-0,scale=.25]{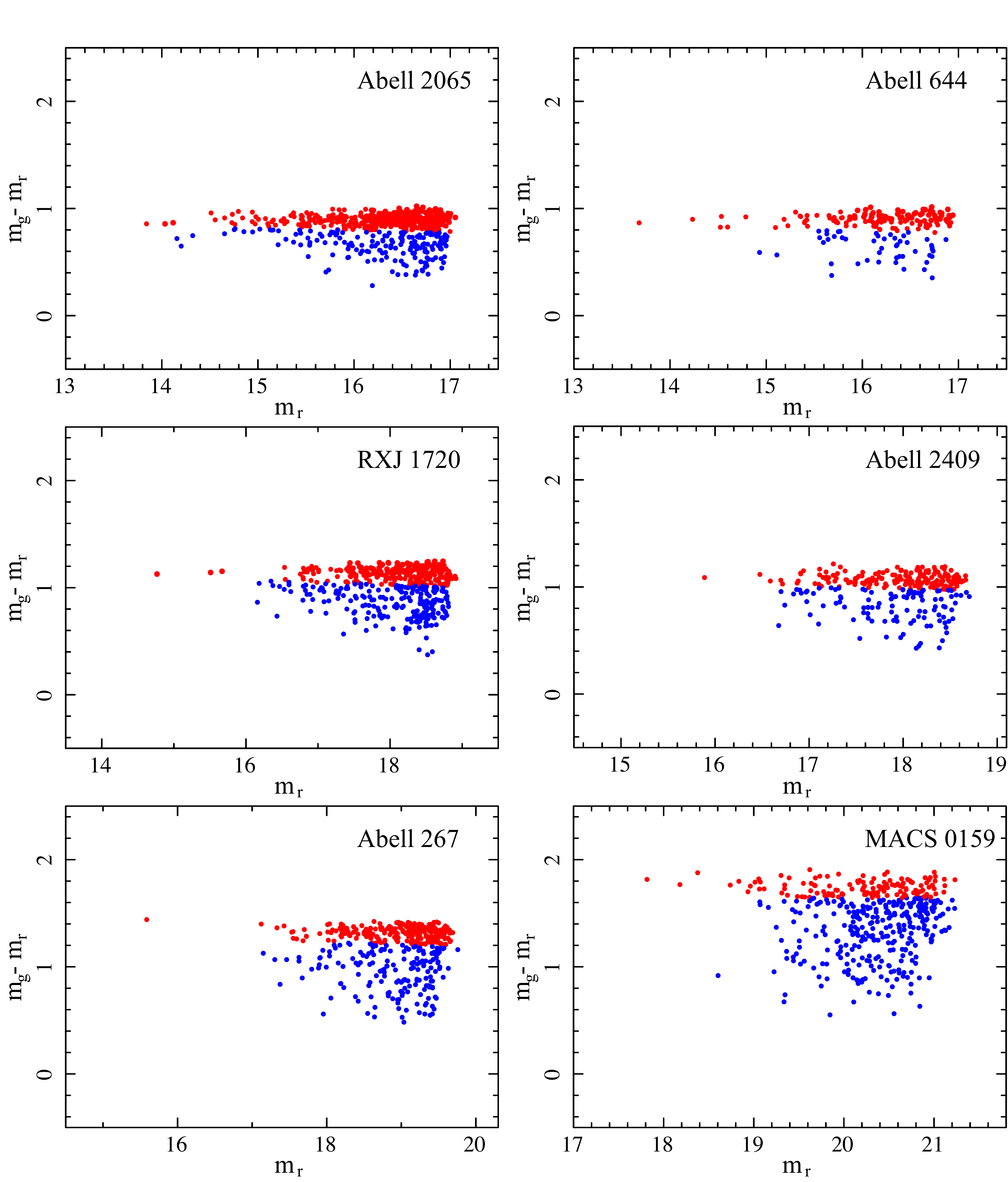}
\caption{Galaxy color-magnitude diagrams ($m_{\rm g} - m_{\rm r}$ vs. $m_{\rm r}$) of six cluster examples. 
The selected red-sequence and blue galaxies are shown in red and blue, respectively. }
\end{center}
\end{figure}

\clearpage

\begin{figure}
\begin{center}
\includegraphics[angle=-0,scale=.4]{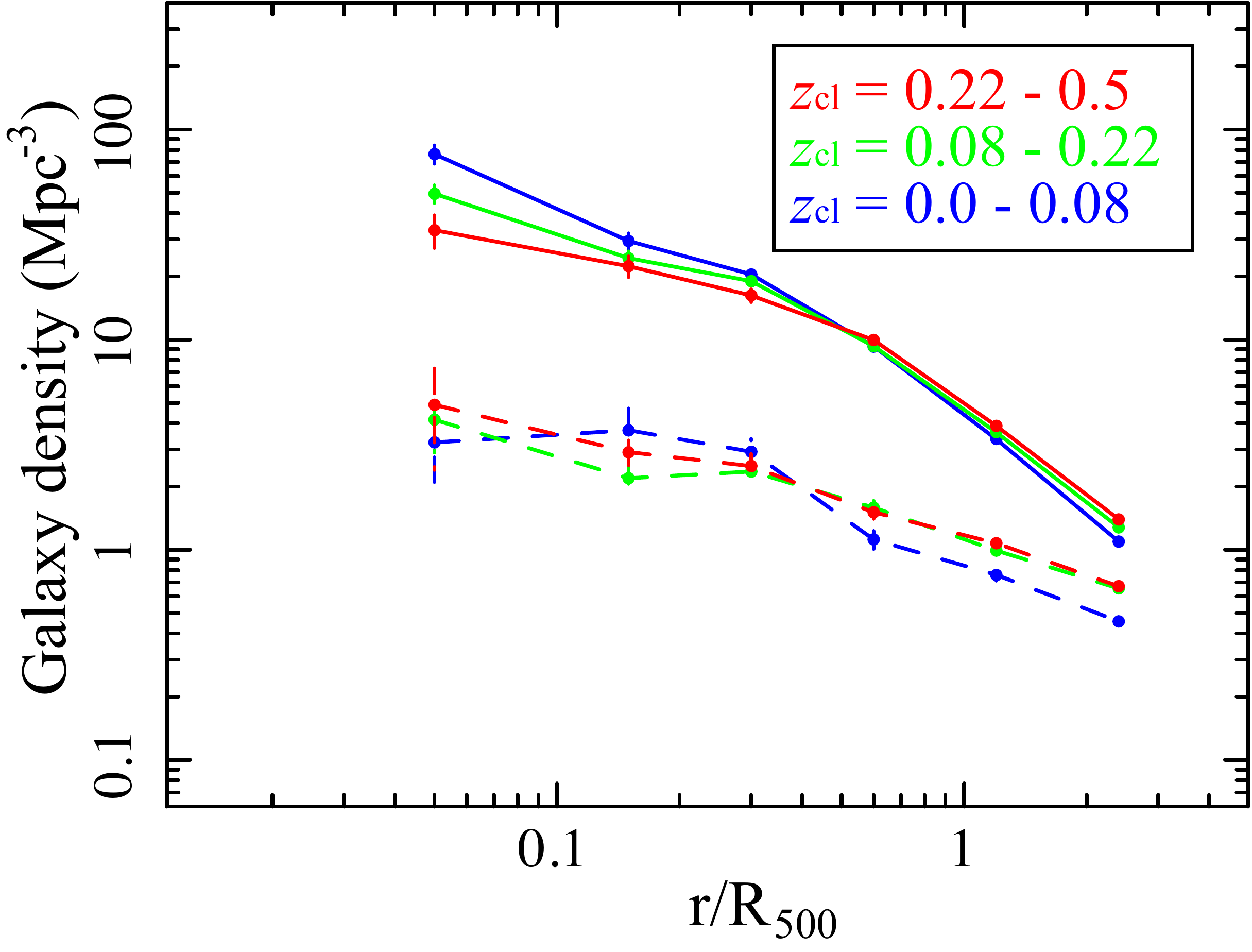}
\caption{Subsample-averaged galaxy density profiles of the color-selected red galaxies (solid lines), 
and the blue ones (dashed line). The blue, green, and red colors stand for subsample L, M, and H, respectively.  }
\end{center}
\end{figure}

\clearpage

\begin{figure}
\begin{center}
\includegraphics[angle=-0,scale=.4]{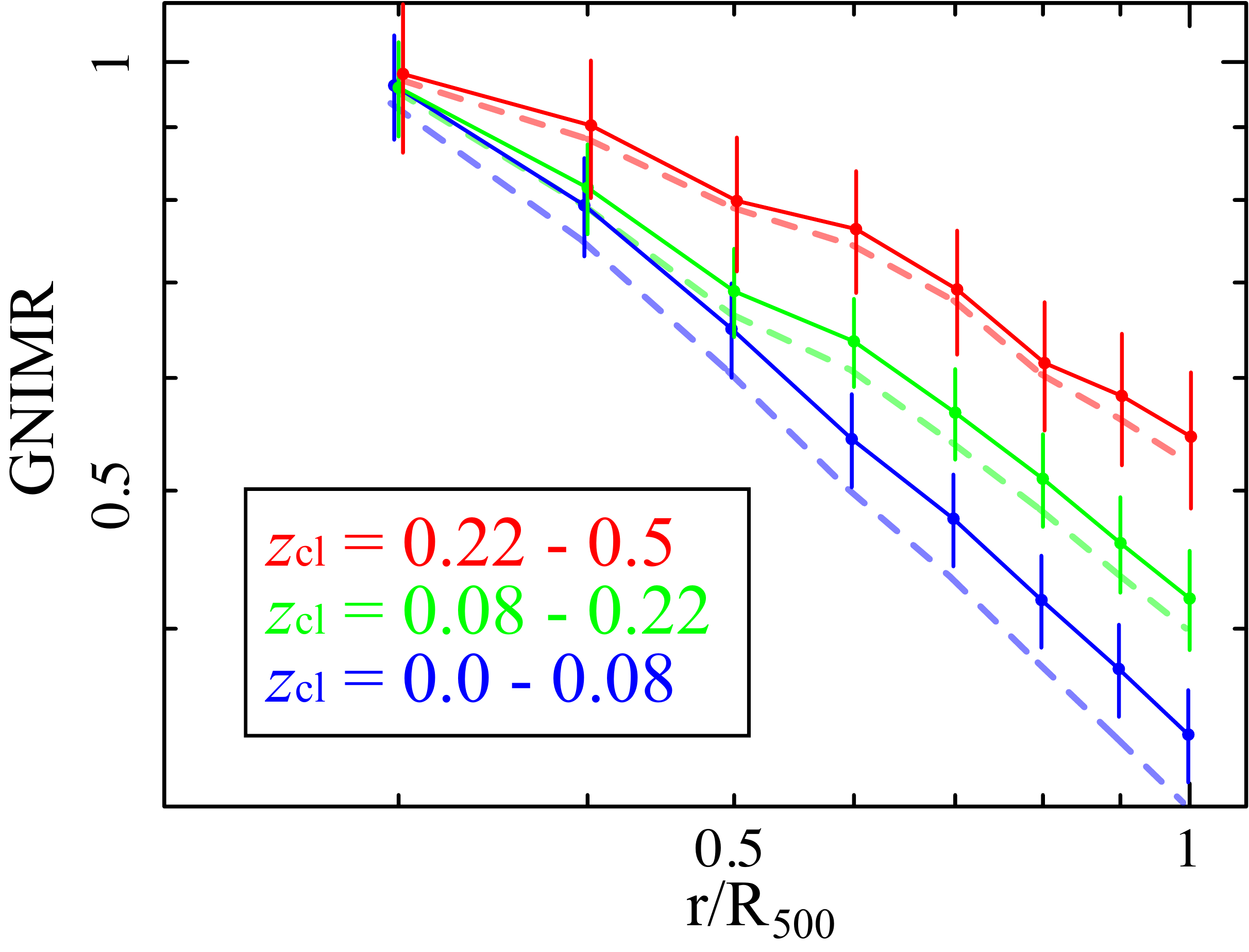}
\caption{Subsample-averaged GNIMR profiles calculated based on low-mass galaxies ($m \leq 1 \times 10^{11} M_{\odot}$;
solid lines), and those with all members (dashed lines). The color of the subsamples is the same as in 
Figure 14b.}
\end{center}
\end{figure}
\clearpage

\begin{figure}
\begin{center}
\includegraphics[angle=-0,scale=.33]{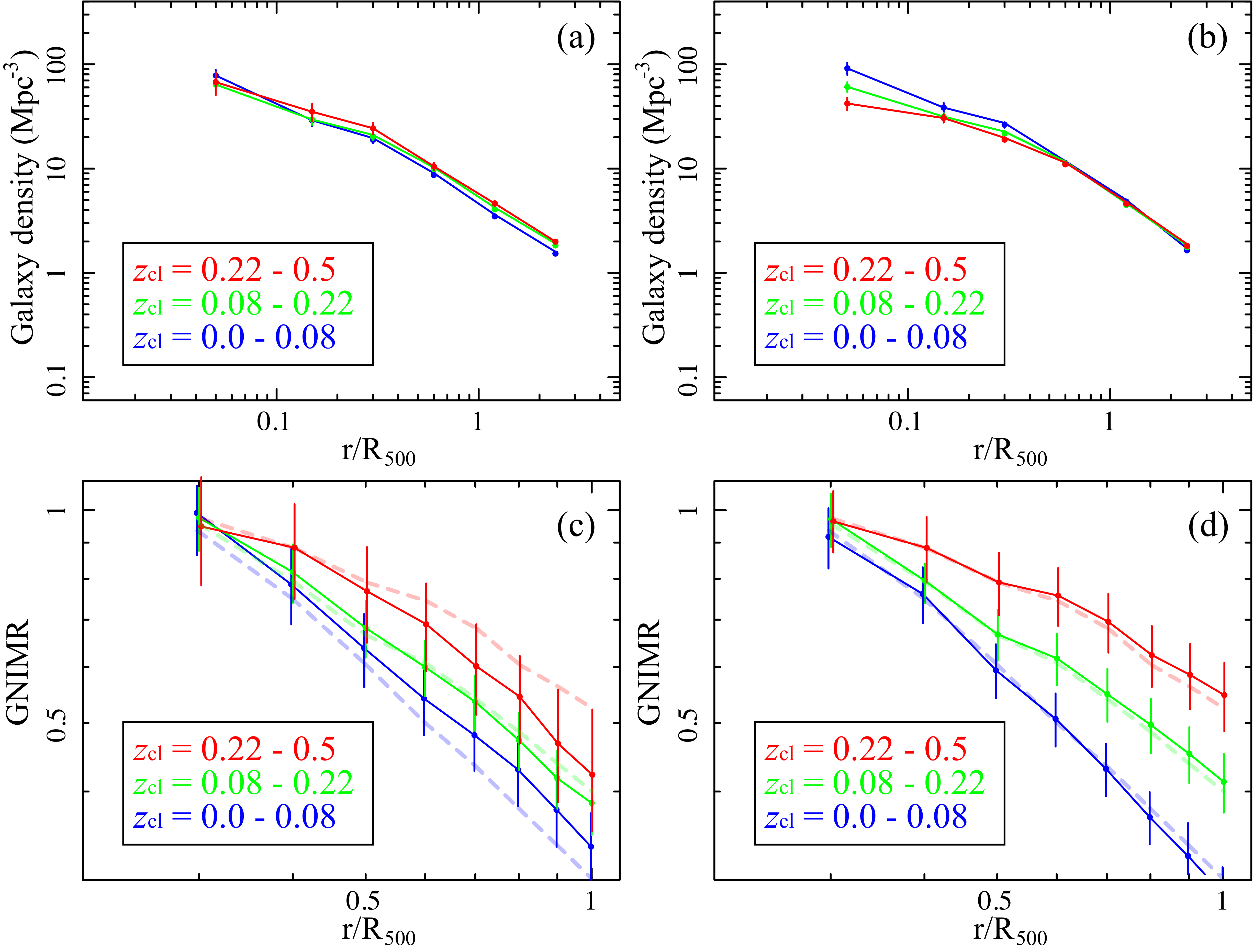}
\caption{Subsample-averaged galaxy number density profiles of clusters with relatively low
(panel a; $\rho_{\rm ICM} \leq 10^{-3}$ cm$^{-3}$) and high (panel b; $\rho_{\rm ICM} > 10^{-3}$ cm$^{-3}$)
ICM densities. The corresponding subsample-averaged GNIMR profiles are given in panels (c) and (d),
where the original GNIMR profiles (Fig. 14b) are plotted as reference in dashed lines. The color of each
subsample is the same as in Figure 14b. }
\end{center}
\end{figure}

\clearpage

\end{document}